\pdfoutput=1
\documentclass[12pt,preprint]{aastex}

\newcommand{\Rmnum}[1]{\expandafter\@slowromancap\romannumeral #1@}

\shorttitle{Solar magnetic cycles}
\shortauthors{Petrie}

\begin{document}


\title{Solar magnetic activity cycles, coronal potential field models and eruption rates}


\author{G.J.D. Petrie}
\affil{National Solar Observatory, Tucson, AZ 85719, USA}



\begin{abstract}
We study the evolution of the observed photospheric magnetic field and the modeled global coronal magnetic field during the past 3 1/2 solar activity cycles observed since the mid-1970s. We use synoptic magnetograms and extrapolated potential-field models based on longitudinal full-disk photospheric magnetograms from the NSO's three magnetographs at Kitt Peak, the Synoptic Optical Long-term Investigations of the Sun (SOLIS) vector spectro-magnetograph (VSM), the spectro-magnetograph and the 512-channel magnetograph instruments, and from the U. Stanford's Wilcox Solar Observatory. The associated multipole field components are used to study the dominant length scales and symmetries of the coronal field. Polar field changes are found to be well correlated with active fields over most of the period studied, except between 2003-6 when the active fields did not produce significant polar field changes. Of the axisymmetric multipoles, only the dipole and octupole follow the poles whereas the higher orders follow the activity cycle. All non-axisymmetric multipole strengths are well correlated with the activity cycle. The tilt of the solar dipole is therefore almost entirely due to active-region fields. The axial dipole and octupole are the largest contributors to the global field except while the polar fields are reversing. This influence of the polar fields extends to modulating eruption rates. According to the Computer Aided CME Tracking (CACTus), Solar Eruptive Event Detection System (SEEDS), and Nobeyama radioheliograph prominence eruption catalogs, the rate of solar eruptions is found to be systematically higher for active years between 2003-2012 than for those between 1997-2002. This behavior appears to be connected with the weakness of the late-cycle 23 polar fields as suggested by Luhmann. We see evidence that the process of cycle 24 field reversal is well advanced at both poles.
\end{abstract}



\section{Introduction}
\label{sect:intro}
\indent 

Since synoptic photospheric magnetogram observations began in the mid-1970s, three full solar activity cycles have been observed and a fourth has begun. As has been widely reported, the recent behavior of the solar magnetic field has stood out as unusual compared to the previous three decades or so of regular synoptic measurements. Since the cycle 23 polar field reversal, the polar fields have been about 40\% weaker than the two previous cycles (Wang et al. 2009, Gopalswamy et al. 2012) and the activity level of cycle 24 has been unusually low, following a very long activity minimum. The weak
polar fields have been linked by Wang et al. (2009) to a 20\% shrinkage in polar
coronal-hole areas and a reduction in the solar-wind mass flux over the poles. 
Low-latitude
coronal holes were larger and more numerous than during the previous minimum
(Lee et al. 2009, de Toma~2011). Large, low-latitude coronal holes were present as late as 2008, finally disappearing in 2009 (de Toma~2011). Meanwhile, the coronal streamer structure and the heliospheric current sheet only became
axisymmetric in the equatorial plane after sunspot numbers fell to unusually
low values in mid-2008 (Wang et al. 2009, Thompson et al.~2011) when the heliospheric flux was at its weakest level
since measurements began in 1967 (Sheeley~2010). The activity level of cycle 24 looks unlikely to reach even the modest heights of cycle 23. At the time of writing, the polar fields are on the point of swapping polarities, consistent with cycle 24 having reached its maximum activity level, and yet the sunspot number is currently little more than half the the typical value recorded during the cycle 23 maximum. In the past, polar field polarity reversals have coincided with activity cycle maxima. According to observations of the ``rush to the poles'' in the Fe~{\sc xiv} corona 1.15 solar radii (Altrock~2013) and prominence eruption activity in microwave brightness observations (Gopalswamy et al. 2012), solar maximum conditions have arrived at the northern hemisphere but not in the south. On the other hand, there seems to be no reason why the sunspot number cannot increase further during this cycle.

Some recent work has been directed towards relating these observed changes in solar magnetism to standard Babcock-Leighton models of the global solar dynamo. Babcock~(1959, 1961) produced a powerful phenomenological description of the solar cycle based on his magnetograph observations. This model was then developed by Leighton~(1964, 1969) into a kinematic model for the transport of photospheric magnetic flux, in which photospheric turbulent diffusion played a key role in decaying active-region fields, causing their leading polarities to interact across the equator and spreading their lagging polarities pole-ward, providing a link between active-region and polar fields. This model was later modified by Wang and Sheeley~(1991) and Wang et al.~(1991) to relate the polar flux distribution to the observed poleward transport of decayed active-region flux by meridional flows. Since then the active-region and polar fields have been coupled in Babcock-Leighton models via a combination of surface diffusion and meridional flow. The recent unusual behavior of the solar field has been modeled by introducing unusual changes in the meridional flow speed in Babcock-Leighton models (e.g., Schrijver and Liu~2008, Wang et al.~2009; Nandy et al.,~2011).

Other work has applied observed solar magnetograms to relate ideas from kinematic dynamo theory to patterns in photospheric magnetic fields and extrapolated coronal models. Using more than 37 years of NSO Kitt Peak and Mt. Wilson 150-foot tower full-disk longitudinal magnetograms, Petrie~(2012) found strong correlation in each hemisphere between poloidal active region fields, high-latitude poleward field surges, and polar field changes, consistent with the relationship between poloidal active region fields and polar field changes in the Babcock-Leighton model. The results also showed that the weak polar fields observed since the maximum of cycle 23 may have been caused by the hemispheric bias of the active region poloidal field component in each hemisphere effectively disappearing after the cycle 23 polar field reversal. DeRosa et al.~(2012) decomposed Wilcox Solar Observatory (WSO) and Michelson Doppler Imager (MDI) field measurements into multipole components and studied the WSO multipoles' evolution over three solar cycles to characterize the dominant large-scale structure of the corona over this time (see also Hoeksema~1984). DeRosa et al. focused on axisymmetric multipole components, and found a coupling between those antisymmetric and symmetric about the equator. They found a corresponding coupling between antisymmetric and symmetric fields in axisymmetric Babcock-Leighton kinematic dynamo solutions that pointed to the necessity of both classes of field for cyclical dynamo behavior to be maintained.

In this paper we seek to extend the above results. We will characterize the relationship between the axisymmetric and non-axisymmetric components of the fields in some detail and with reference to the sunspot number and polar field measurements from NSO and WSO. Separating the fields into classes symmetric and asymmetric about the rotation axis and symmetric and anti-symmetric about the equator, we will determine which classes follow the sunspot activity cycle and which are dominated by the polar fields. The evolving relationship between the active and polar fields over time will be investigated and unusual behavior reported. We will also determine whether the field evolution can shed light on recent unexpected changes in activity indices, in particular the coronal mass ejection (CME) rate and the prominence eruption (PE) rate. The relationship between CMEs and PEs has been extensively studied in the past. Studying the association rate, relative timing, and spatial correspondence between PEs and CMEs, Gopalswamy et al.~(2003) found that most (72\%) PEs were clearly associated with CMEs and that during the solar minimum, the central position angle of the CMEs tends to be offset closer to the equator compared to the PEs but not during solar maximum (see also Gopalswamy et al.~2012). Shimojo et al.~(2006) found that the prominence eruption rate follows the sunspot number, with differences between their peak times, and describes a butterfly pattern but with some hemispheric asymmetry. Luhmann et al.~(2012) have noted that the cycle 24 CME rate is comparable to the cycle 23 rate even though the cycle 24 sunspot number is significantly weaker. They suggested that this phenomenon might be related to the weakness of the polar fields since the cycle 23 polarity reversal. We will explore this suggestion using the magnetic field data and solar eruption statistics from three well-known public catalogs.

The paper is organized as follows. Section~\ref{sect:data} describes the various data sets analyzed in the paper. In Section~\ref{sect:activepolar} the evolution of the active and polar fields, and the axisymmetric and non-axisymmetric multipoles, are described in detail. The relationship between these fields and the CME and PE rates is explored in Section~\ref{sect:cmes}. We conclude in Section~\ref{sect:conclusion}.

\section{Data}
\label{sect:data}

We analyze synoptic maps for the radial photospheric magnetic field component from three NSO Kitt Peak magnetographs that have been observing the full-disk longitudinal photospheric field daily, weather permitting, since 1974: the Synoptic Optical Long-term Investigations of the Sun (SOLIS) vector-spectromagnetograph (VSM), that has been observing the 630.2~nm Fe~{\sc i} spectral line since August 2003 (Keller et al.,~2003); the spectro-magnetograph (SPMG), that observed the 868.8~nm  Fe~{\sc i} spectral line from November 1992 until September 2003 (Jones et al.,~1992); and the 512-channel instrument, that observed the 868.8~nm  Fe~{\sc i} spectral line from February 1974 until April 1993 (Livingston et al.,~1976). The earliest synoptic map covers Carrington rotation 1626 in March 1975. Together the data set covers cycles 21-23 in their entirety, as well as the end of cycle 20, and cycle 24 up to the present. The sky images were taken at $1^{\prime\prime}$ pixel$^{-1}$ spatial resolution and $360\times 180$-pixel full-surface synoptic maps for the radial field were constructed in longitude-sin(latitude) coordinates by software pipelines running at NSO. 

We also analyze time series of synoptic photospheric radial magnetic field data derived from measurements of the line-of-sight field taken at WSO using the Fe~{\sc i} spectral  line at 525.02~nm (Scherrer et al.~1977). The WSO magnetograph scans the solar image with a $175^{\prime\prime}\times 175^{\prime\prime}$ square aperture. The WSO synoptic maps cover about 36 1/2 years from May 1976 (Carrington rotation 1642) until the present. The data are analyzed here in the form of spherical harmonic coefficients calculated rotation by rotation at WSO and made available online\footnote{http://wso.stanford.edu}, as well as estimates of the polar field strengths taken every 10-days over a 30-day window\footnote{http://wso.stanford.edu/Polar.html}.

The NSO data clearly have much higher spatial resolution than the WSO data but a major advantage of the WSO data is that the WSO telescope has been operating continuously since 1976 with no instrument changes. The NSO/KP data set, deriving from three different instruments, has to be cross-calibrated. The analysis is complicated by the fact that the NSO sensitivity and saturation levels vary significantly among the three NSO instruments, and so the results must be interpreted with care. The WSO data can be analyzed without the complication of cross-calibration issues. However, there are known saturation errors in the WSO data generally. Following DeRosa et al.~(2012) we do not correct the WSO data for known saturation effects that cause the fields to be underestimated by a factor of about 1.8 (Svalgaard et al.~1978).

In a later section we will investigate the connection between the active and polar fields on the one hand and the rate of solar eruptions on the other. The Computer Aided CME Tracking project (CACTus, Robbrecht et al.~2009) automatically detects coronal mass ejections (CMEs) in image sequences from NASA's Large Angle and Spectrometric Coronagraph (LASCO) without human intervention\footnote{http://sidc.oma.be/CACTus/catalog/LASCO}. Compared to catalogs assembled manually by human operators, these automatic CME detections might be more objective as the detection criterion is written explicitly in a program but the lack of manual filtering also means that the catalog needs to be treated with some caution. The detection method and the catalog are described and analyzed in detail by Robbrecht et al.~(2009), who found the statistical properties of the derived CME rates to be encouraging. Until 2010 the algorithm ran on level 0 images from the LASCO C2 and C3 instruments and since 2010 quick-look images from these instruments have been used.

The Solar Eruptive Event Detection System (SEEDS) project at George Mason University's Space Weather Laboratory also runs an automated algorithm to detect, classify and characterize eruptive events including CMEs (Olmedo et al.~2008). A catalog of CMEs covering the LASCO data set has been collected\footnote{http://spaceweather.gmu.edu/seeds/lasco.php}. We will determine whether the CACTus and SEEDS statistics give consistent results.

We also use the catalog of limb prominence eruption (PE) events\footnote{
http://solar.nro.nao.ac.jp/norh/html/prominence/} derived from microwave imaging observations at 17 GHz made by the Nobeyama radioheliograph (Nakajima et al.~1994). The Nobeyama radioheliograph is a radio interferometer offering full-disk coverage with $10^{\prime\prime}$ resolution at 17GHz and temporal resolution of 1 second.

\section{Active and Polar Fields}
\label{sect:activepolar}

\subsection{The butterfly diagram, sunspot number and polar fields}
\label{s:butterfly}

\begin{figure} 
\begin{center}
\resizebox{0.99\textwidth}{!}{\includegraphics*{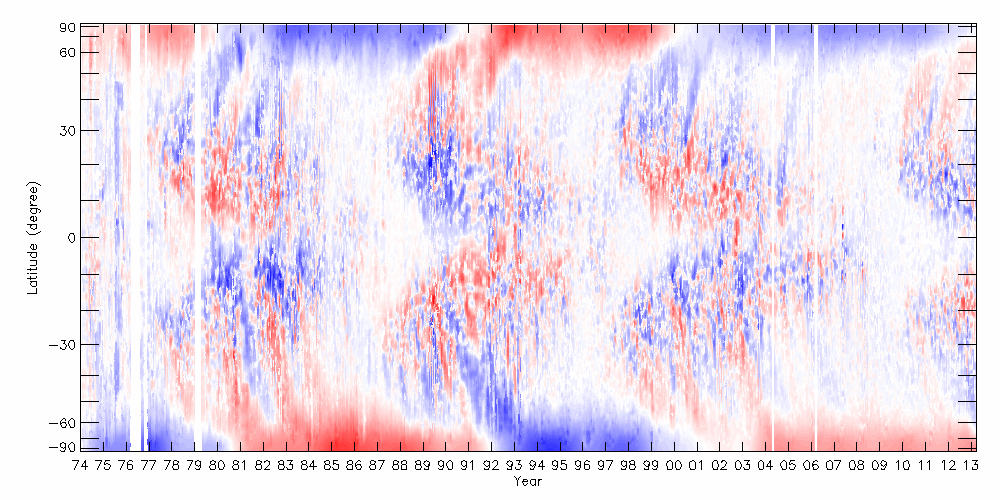}}
\end{center}
\caption{Butterfly diagram based on Kitt Peak data, summarizing the photospheric radial field distributions derived from the longitudinal photospheric field measurements. Each pixel is colored to represent the average field strength at each time and latitude. Red/blue represents positive/negative flux, with the color scale saturated at $\pm 20$~G.}
\label{fig:butterfly}
\end{figure}

\begin{figure}[h]
\begin{center}
\resizebox{0.69\hsize}{!}{\includegraphics*{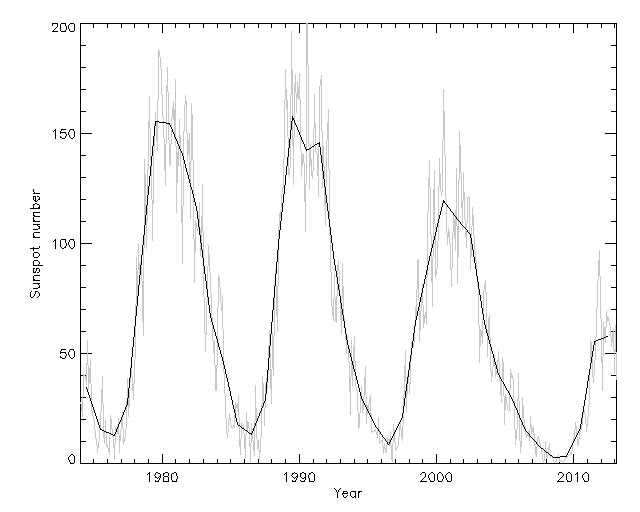}}
\end{center}
\caption{The monthly sunspot number from 1974 to 2013. The grey curve shows the monthly sunspot number and the black curve the annual averages.}
\label{fig:ssn}
\end{figure}

\begin{figure}[h]
\begin{center}
\resizebox{0.69\hsize}{!}{\includegraphics*{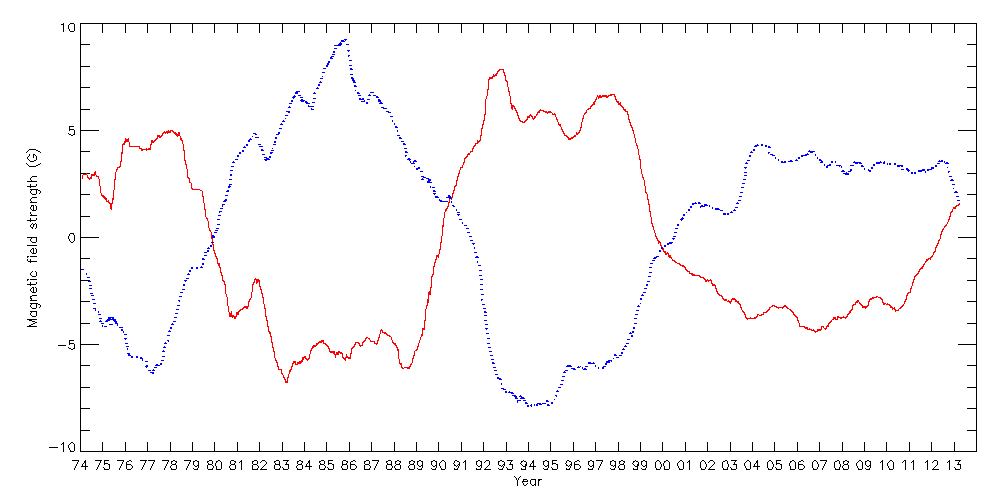}}
\resizebox{0.69\hsize}{!}{\includegraphics*{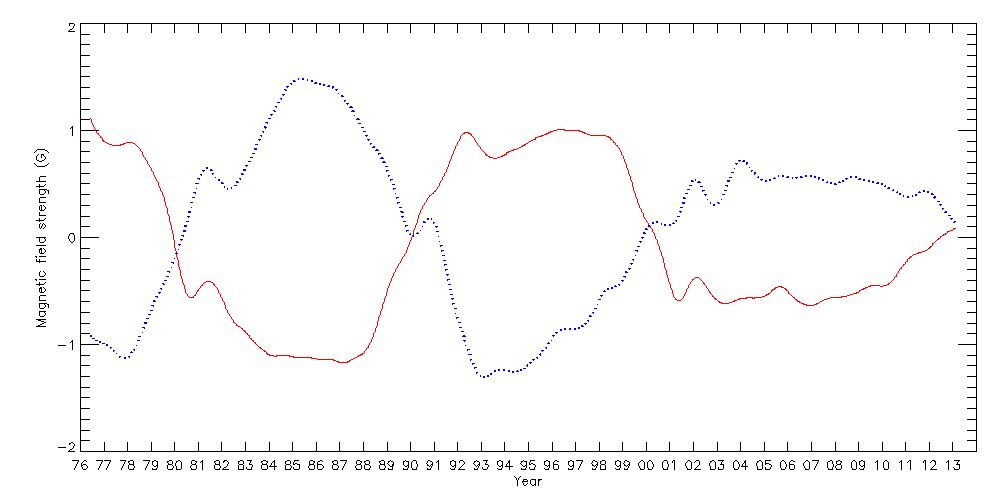}}
\end{center}
\caption{10-day averages of the north (red solid lines) and south (blue dotted lines) measured at Kitt Peak (top) and Wilcox (bottom) are plotted against time. The Kitt Peak data are for the radial field component and derive from heliographic latitudes between about latitudes ranging from about $\pm 63^{\circ}$ to about $\pm 70^{\circ}$. The Wilcox measurements are for the line-of-sight field component and come from the 3' apertures nearest the poles, which cover between about $\pm 55^{\circ}$ and the poles.}
\label{fig:poles}
\end{figure}

Figure~\ref{fig:butterfly} shows the NSO/KP magnetic butterfly diagram, updated from Figure~1 of Petrie~(2012) to early 2013 and smoothed using a 27-day boxcar filter. This plot was derived using the full-disk sky images collected at NSO/KP since February 1974. The diagram clearly shows the cyclical behavior of the active and polar fields. Features of the Babcock-Leighton model of the solar cycle are clearly evident. The active fields begin each cycle emerging at latitudes around $\pm 30^{\circ}$ and subsequently emerge at progressively lower latitudes on average, creating the wings of the distinctive butterfly patterns. The diagram also shows the change of polarity of the polar fields once each cycle, coinciding with activity maximum (Babcock~1959). Most of the flux that emerges in active regions cancels with flux of opposite polarity but a proportion survives as weak flux that is carried poleward by the meridional flow. This poleward drift of the weak, decayed magnetic flux appears as plumes of one dominant polarity at high latitudes, between about $40^{\circ}$ and $65^{\circ}$, corresponding to changes in the polar field. In each hemisphere during each cycle the plumes generally have a strong tendency to take the polarity of the lagging polarity of the active regions. These patterns reflect the well-known facts that magnetic bipoles are typically asymmetric and tilted so that the leading polarity is stronger, more compact and slightly closer to the equator than the following polarity (Hale et al.~1919), and the vast majority of bipoles in each hemisphere have the same leading polarity, the leading polarities are opposite in the two hemispheres during any given cycle, and all polarity patterns reverse each cycle (Hale and Nicholson 1925). Thus the active and polar fields appear to be linked together in a single magnetic cycle~(Babcock~1961). Leighton~(1964, 1969) showed that the active-region and polar fields could be coupled by the effects of photospheric diffusion, and Wang \& Sheeley~(1991) and Wang et al.~(1991) demonstrated that systematic pole-ward meridional flows likely play a key role in coupling the two classes of field in a global activity cycle. Figure~\ref{fig:butterfly} clearly shows that the recent cycle 23 minimum was unusually long and quiet. Figure~\ref{fig:butterfly} also shows that cycle 23's polar fields also stand out as being weaker than the polar fields for any other cycle, and the plumes of decayed active-region flux were generally weak and of unusually mixed polarity following the cycle 23 polar reversal. The relationship between these two latter observations was explored by Petrie~(2012), as is the hemispheric asymmetry of the activity.

Figure~\ref{fig:ssn} shows the variation of the sunspot number\footnote{http://www.ngdc.noaa.gov/stp/solar/ssndata.html} over the same period. The sunspot maxima/minima coincide with the activity maxima/minima in Figure~\ref{fig:butterfly}, including the unusually long minimum of cycle 23 and the unusually low activity level of cycle 24. The sunspot number does not represent all active-region fields, especially during a weak cycle such as the present one, but it is a standard index of solar activity and is generally found to correlate well with other activity indices.

Figure~\ref{fig:poles} shows estimates of the average field strengths at polar latitudes based on measurements from the two observatories. At WSO the north and south polar line-of-sight field strength is measured daily in the $3^{\prime}$ apertures nearest the poles, north and south. Every 10 days an average is derived for each pole from measurements in a 30-day centered window. The NSO/KP estimates derive from the butterfly map shown in Figure~\ref{fig:butterfly} and represent the radial field component. This is part of the reason why the NSO measurements are so much stronger than the WSO measurements. There are also well documented calibration issues with the WSO data (Svalgaard et al.~1978) causing the fields to be underestimated by a factor of 1.8. Setting these differences aside, the time series shown in the two panels of Figure~\ref{fig:poles} are clearly well correlated with each other. They agree, for example, that the polar fields were only about 60\% as strong during the cycle 23 minimum compared to the cycle 22 minimum. The two plots also agree that the south polar field steadily decreased in strength beginning early in 2010 reversed polarity around mid-2012, and that the recent change in the north polar field has been more abrupt. The difference in latitude range between the NSO data (between about $\pm 63^{\circ}$ to about $\pm 70^{\circ}$) and the WSO data (between about $\pm 55^{\circ}$ and the poles) in Figure~\ref{fig:poles} cause the change in the north polar field to appear more gradual and to begin earlier in the WSO plot than in the NSO plot. However, both the NSO and WSO north pole curves agree that the north polar field is rapidly declining towards zero and Figure~\ref{fig:butterfly} shows that the last remains of a positive-polarity polar cap survive south of $-70^{\circ}$.

\subsection{Finite-difference potential-field source-surface models for the corona}
\label{sect:fdpfss}

The response of the coronal magnetic field to the photospheric activity patterns can be diagnosed in a simple way by calculating extrapolated potential-field source-surface (PFSS, Altschuler and Newkirk~1969, Schatten et al.~1969) models using the NSO and WSO synoptic maps as lower boundary conditions. Low in the corona, the magnetic field is sufficiently dominant over the plasma forces that a force-free field approximation is generally applicable. Moreover, for large-scale coronal structure the effects of force-free electric currents, which are inversely proportional to length scale, may be neglected. A potential-field model for the corona must satisfy,

\begin{eqnarray}
{\bf\nabla}\times{\bf B} & = & 0,\label{eq:currentfree}\\
{\bf\nabla}\cdot{\bf B} & = & 0.\label{eq:solenoidal}
\end{eqnarray}

\noindent The solution of Equation~(\ref{eq:currentfree}) can be represented as the gradient of a scalar potential, ${\bf B} = -{\bf\nabla}\psi$. Equation~(\ref{eq:solenoidal}) therefore becomes

\begin{equation}
{\bf\nabla}^2\psi = 0.\label{eq:Laplace}
\end{equation}

We identify the lower boundary of the model with the solar photosphere at $r=R$ where $R$ is the solar radius. We use the photospheric radial field maps from NSO to fix $B_r = -\partial\psi /\partial r$ at the lower boundary $r=R$. The use of radial field data, derived from longitudinal field measurements assuming the magnetic vector to be approximately radial in general, has been found to be more successful in reconstructing coronal magnetic structure than the direct application of longitudinal measurements as boundary data (Wang and Sheeley~1992). A synoptic map construction method that derives the radial component from longitudinal measurements without assuming the boundary measurements to be approximately radial has been developed and applied to chromospheric data by Jin et al.~(2013) with promising results. Here we adopt the standard photospheric synoptic maps because measured photospheric fields are found to be approximately radial in general (Svalgaard et al.~1978, Petrie and Patrikeeva~2009, Gosain and Pevtsov~2012) and the resulting models remain competitive for reasons explained by Wang and Sheeley~(1992).

Above some height in the corona, usually estimated to be between 1.5 and 3.5 solar radii, the magnetic field is dominated by the thermal pressure and inertial force of the expanding solar wind. To mimic the effects of the solar wind expansion on the field, we introduce an upper boundary at $r=R_s > R$, and force the field to be radial on this boundary by setting $\psi =0$ there, following Altschuler and Newkirk~(1969), Schatten et al.~(1969) and many subsequent authors. The usual value for $R_s$ is $R_s=2.5R$ although different choices of $R_s$ lead to more successful reconstructions of coronal structure during different phases of the solar cycle (Lee et al.~2011). For simplicity we adopt the standard value $R_s=2.5R$ in this work. With these two boundary conditions, the potential field model can be fully determined in the domain $R\le r\le R_s$.

We use the National Center for Atmospheric Research's (NCAR) MUDPACK\footnote{http://www2.cisl.ucar.edu/resources/legacy/mudpack} package to solve Equation~(\ref{eq:currentfree}) numerically subject to the above boundary conditions. Although Equation~(\ref{eq:currentfree}) can be solved analytically, and has been so treated for several decades, we adopt a finite-difference approach in this subsection to avoid some problems associated with the usual approach based on spherical harmonics (T\'{o}th et al.~2011). MUDPACK (Adams~1989) is a package for efficiently solving linear elliptic Partial Differential Equations (PDEs), both separable and non-separable, using multigrid iteration with subgrid refinement procedures. By iterating and transferring approximations and corrections at subgrid levels, a good initial guess and rapid convergence at the fine grid level can be achieved. Multigrid iteration requires less storage and computation than direct methods for non-separable elliptic PDEs and is competitive with direct methods such as cyclic reduction for separable equations. In particular, three-dimensional problems can often be handled at reasonable computational cost. We use the MUDPACK package here to solve Equation~(\ref{eq:Laplace}) in its non-separable form in spherical coordinates, subject to the boundary conditions described above. Jiang and Feng~(2012) have recently presented a high-speed combined spectral/finite-difference PFSS solution method using the related NCAR FISHPACK package.

\begin{figure}[h]
\begin{center}
\resizebox{0.31\hsize}{!}{\includegraphics*{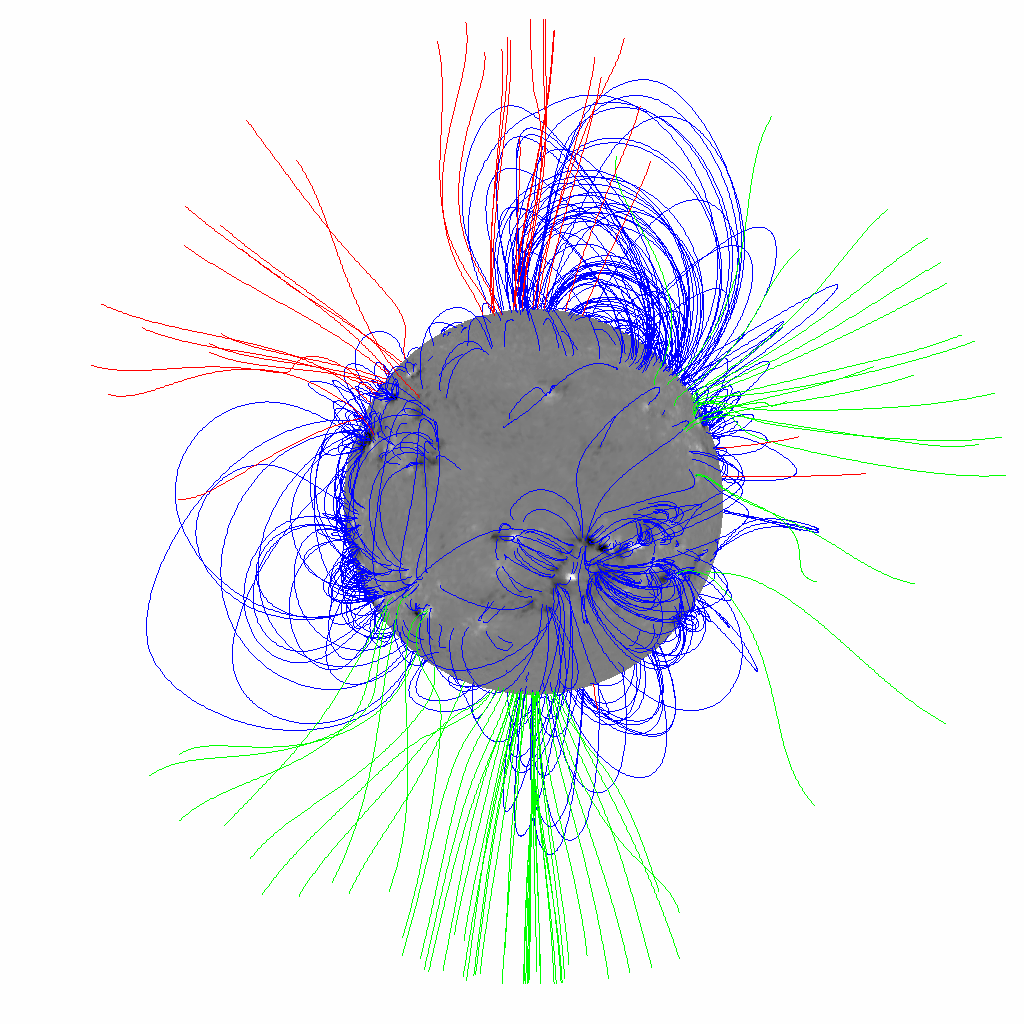}}
\resizebox{0.31\hsize}{!}{\includegraphics*{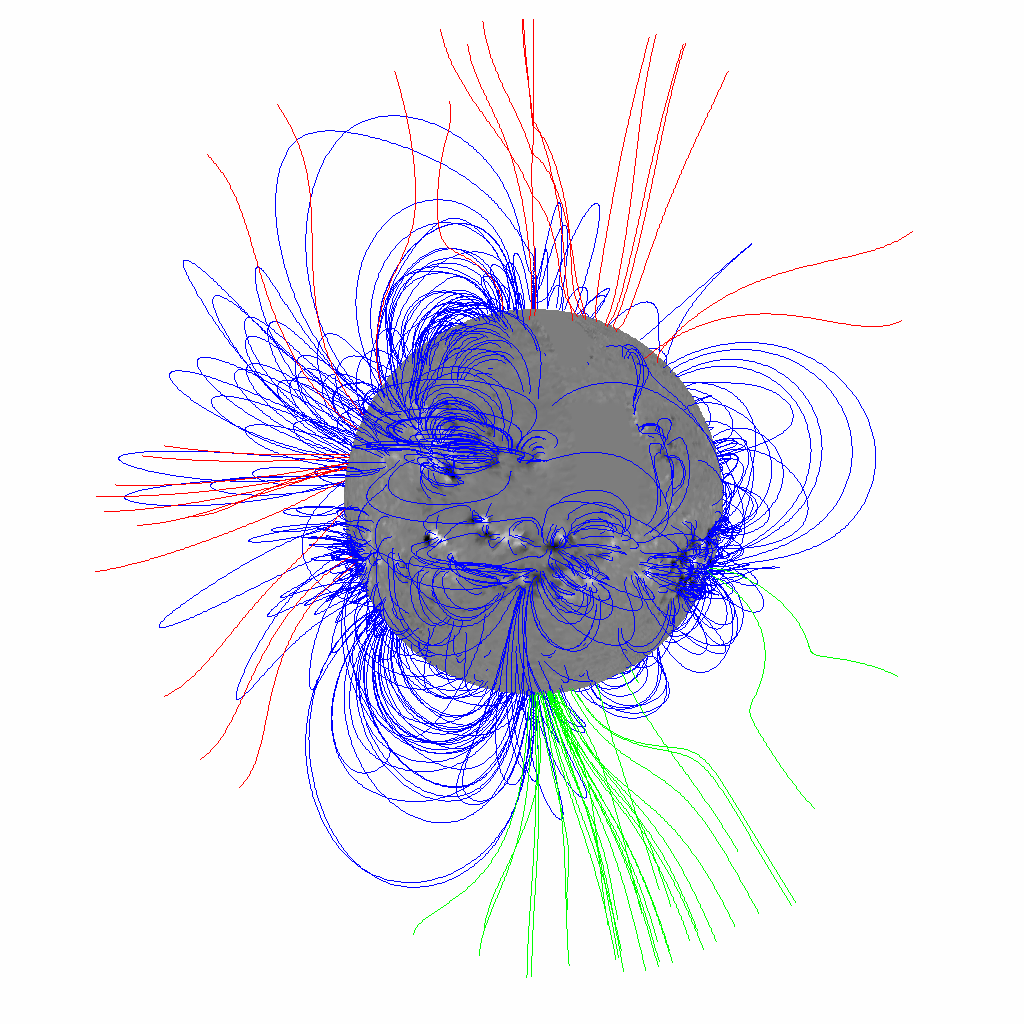}}
\resizebox{0.31\hsize}{!}{\includegraphics*{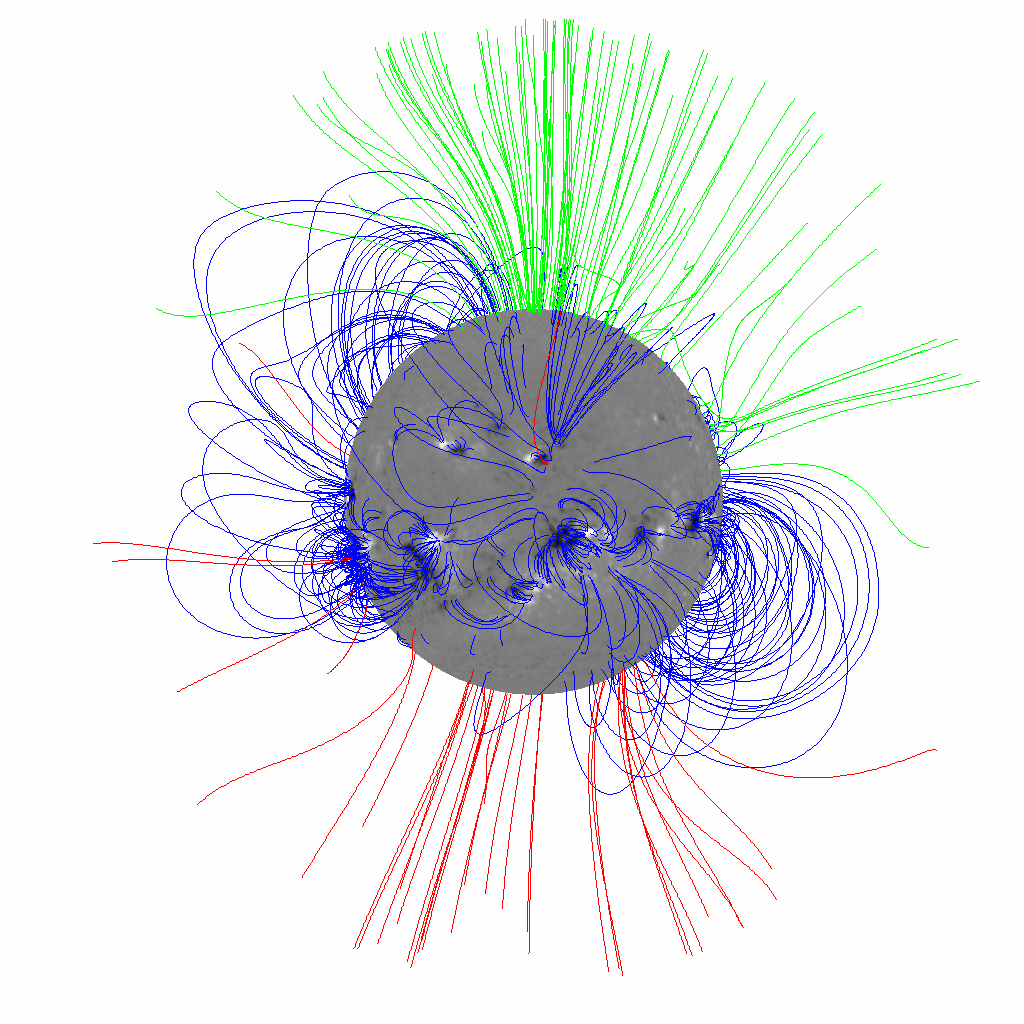}}
\resizebox{0.31\hsize}{!}{\includegraphics*{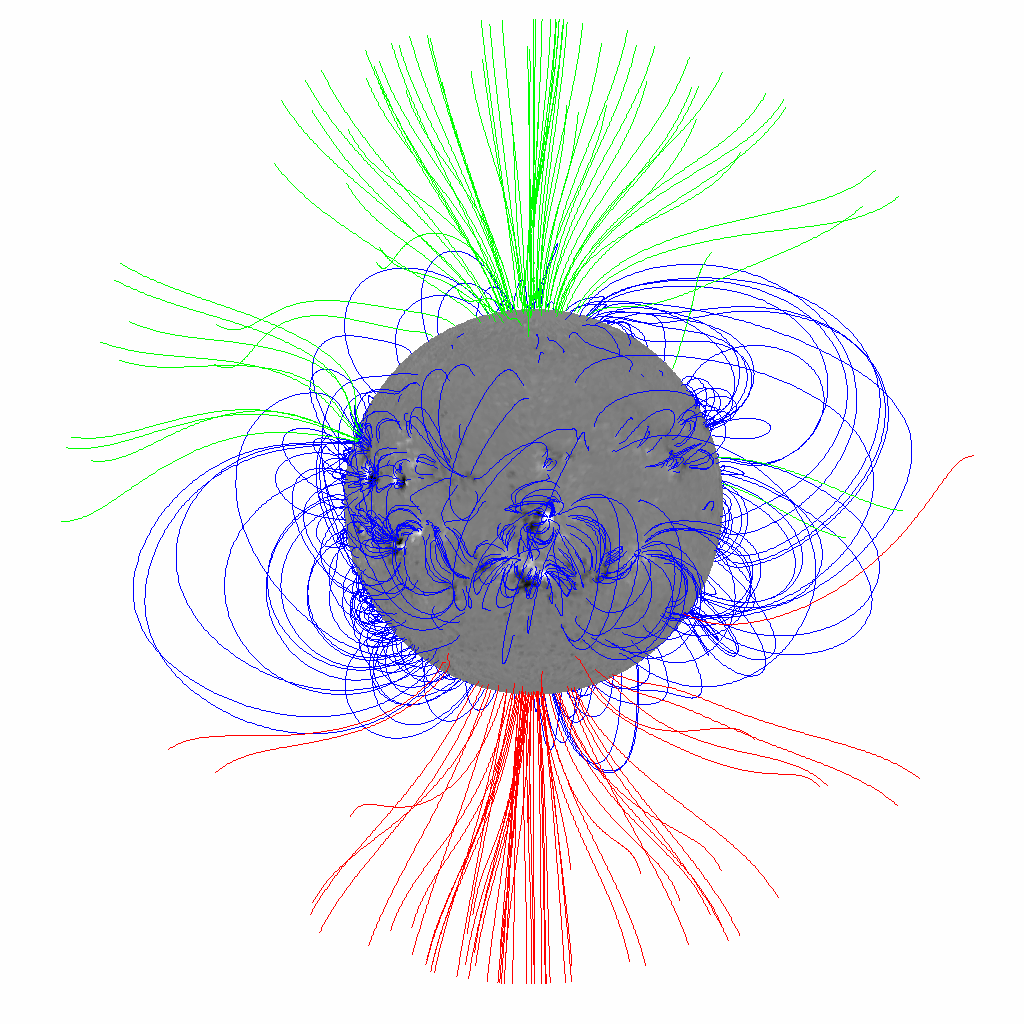}}
\resizebox{0.31\hsize}{!}{\includegraphics*{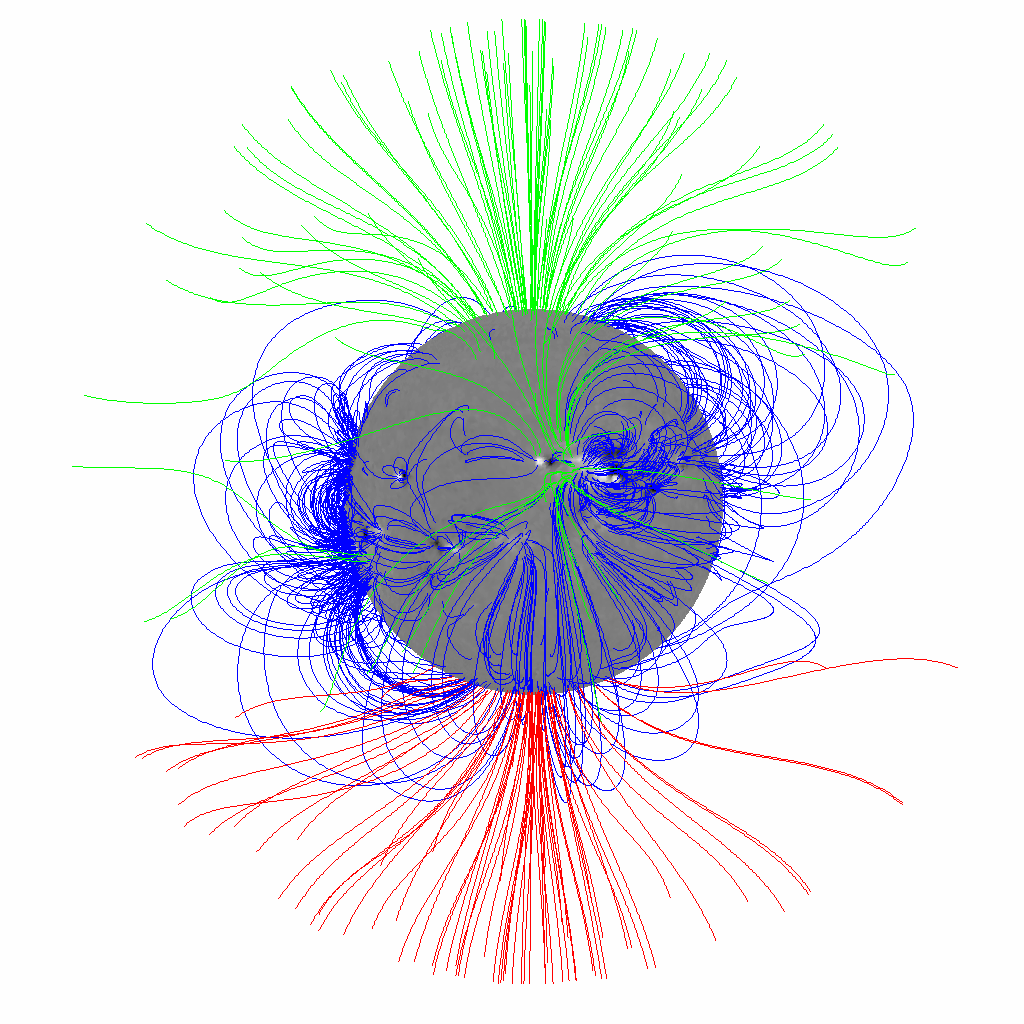}}
\resizebox{0.31\hsize}{!}{\includegraphics*{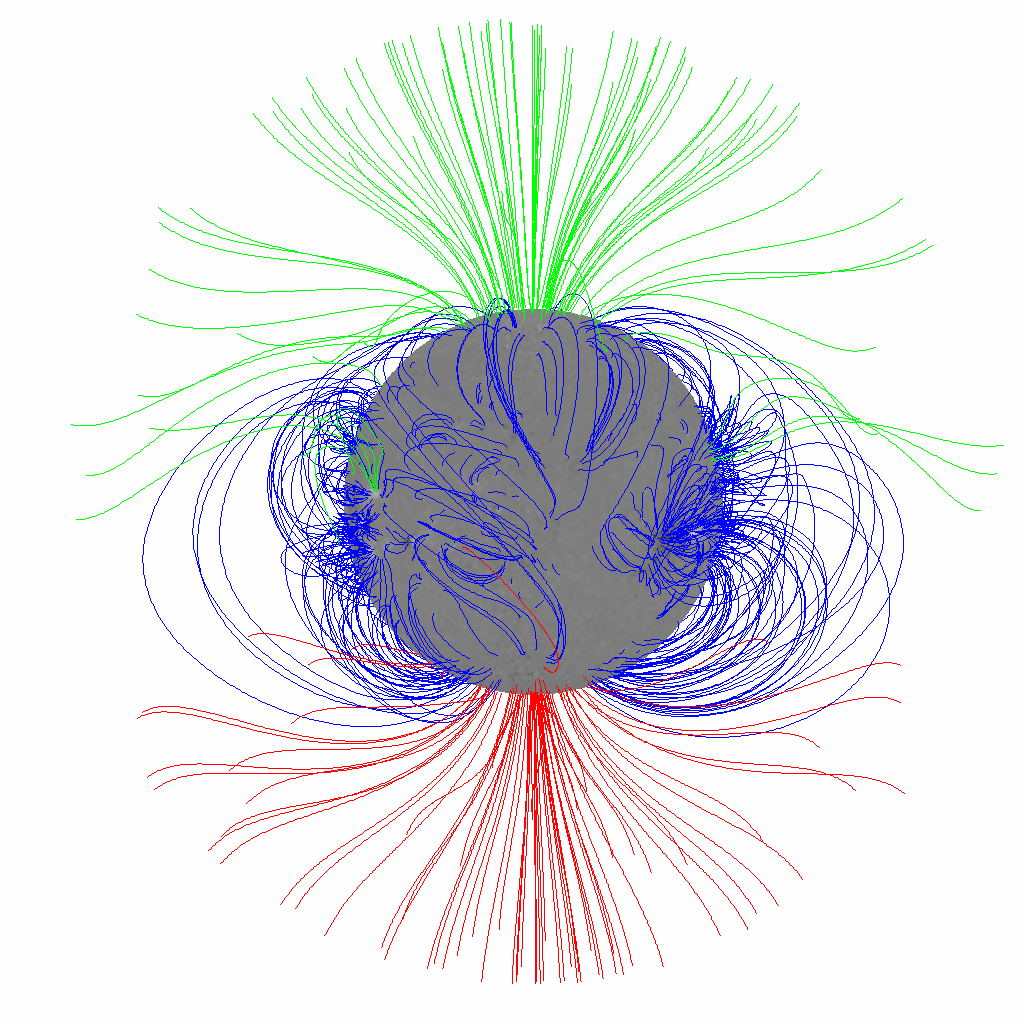}}
\resizebox{0.31\hsize}{!}{\includegraphics*{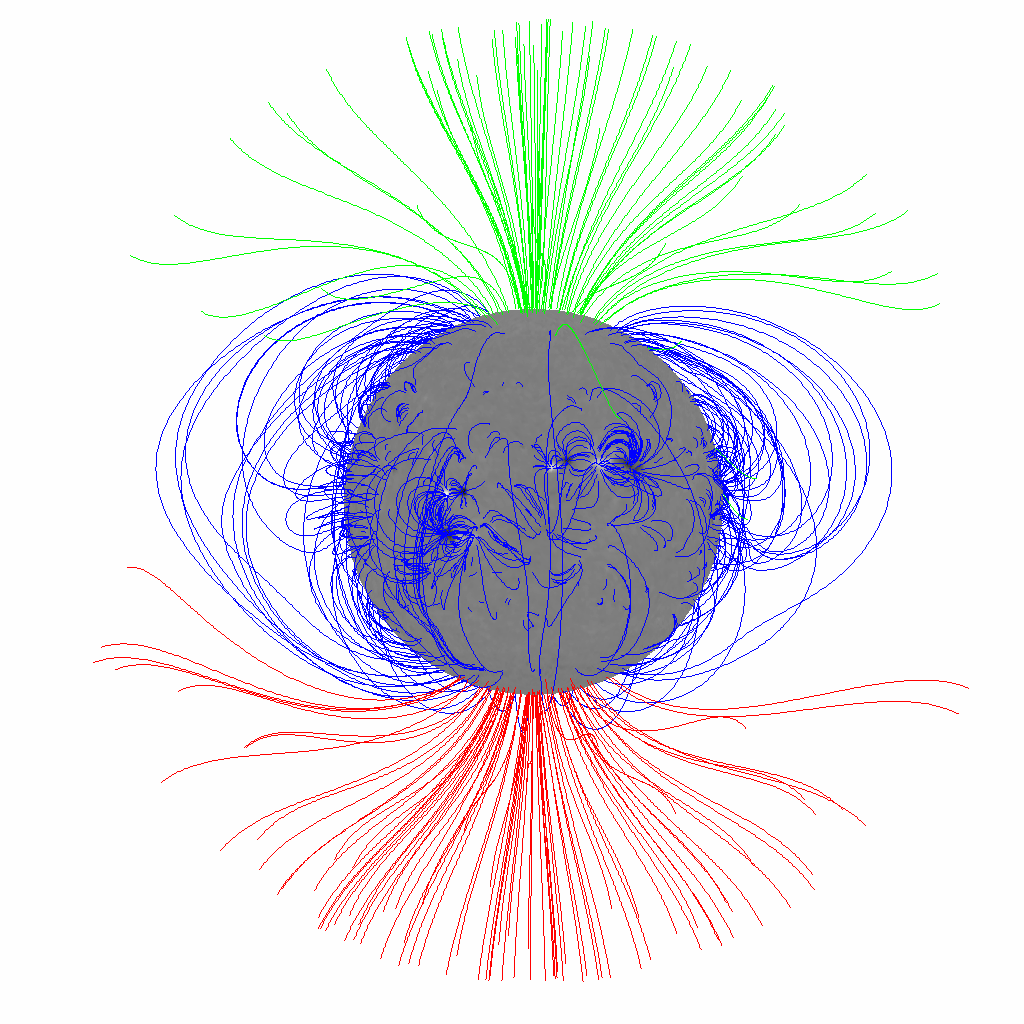}}
\resizebox{0.31\hsize}{!}{\includegraphics*{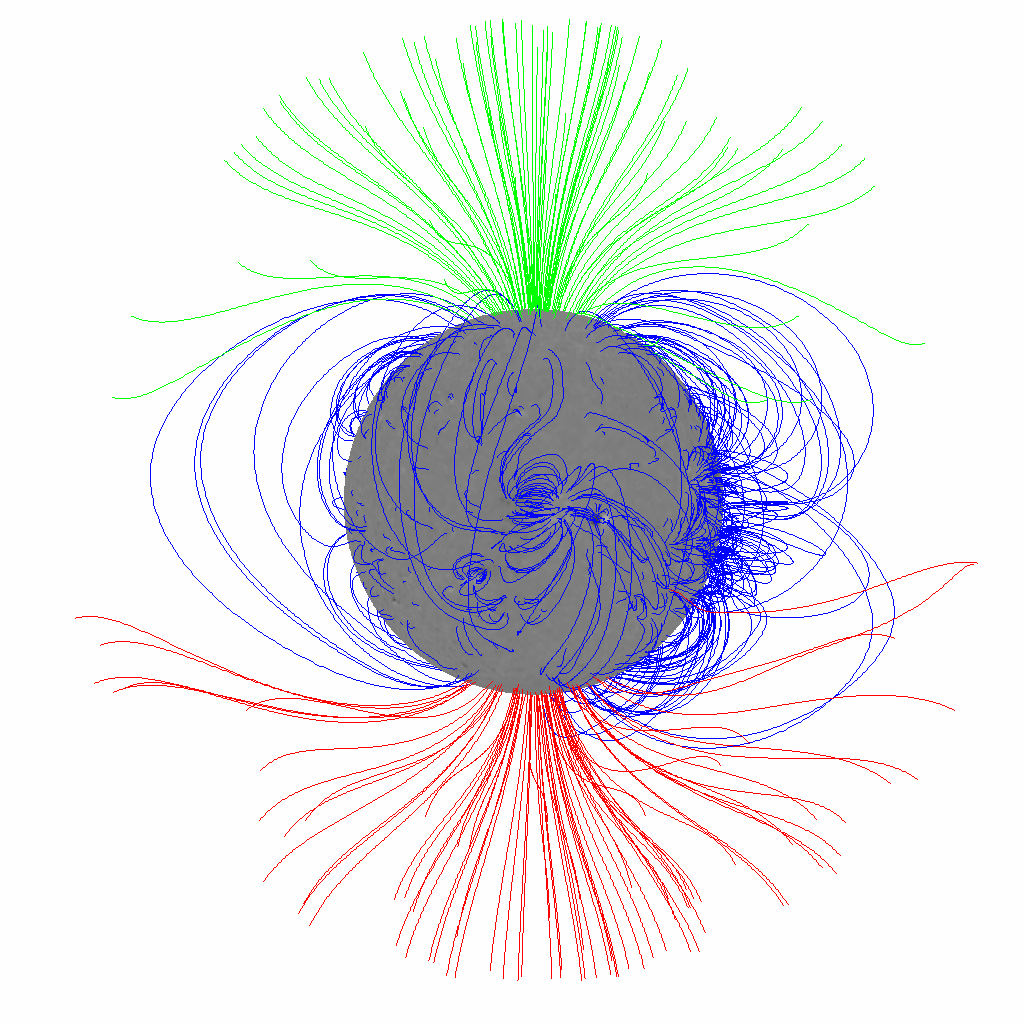}}
\resizebox{0.31\hsize}{!}{\includegraphics*{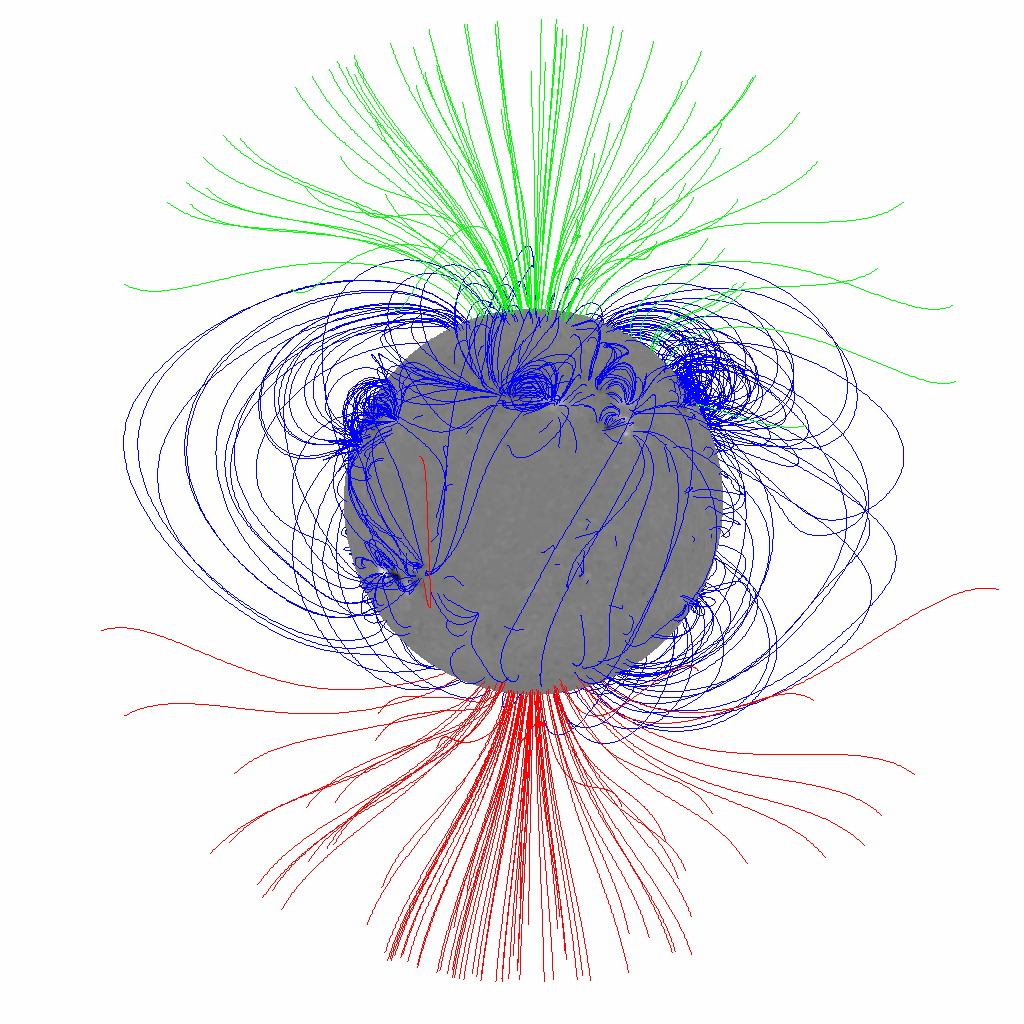}}
\resizebox{0.31\hsize}{!}{\includegraphics*{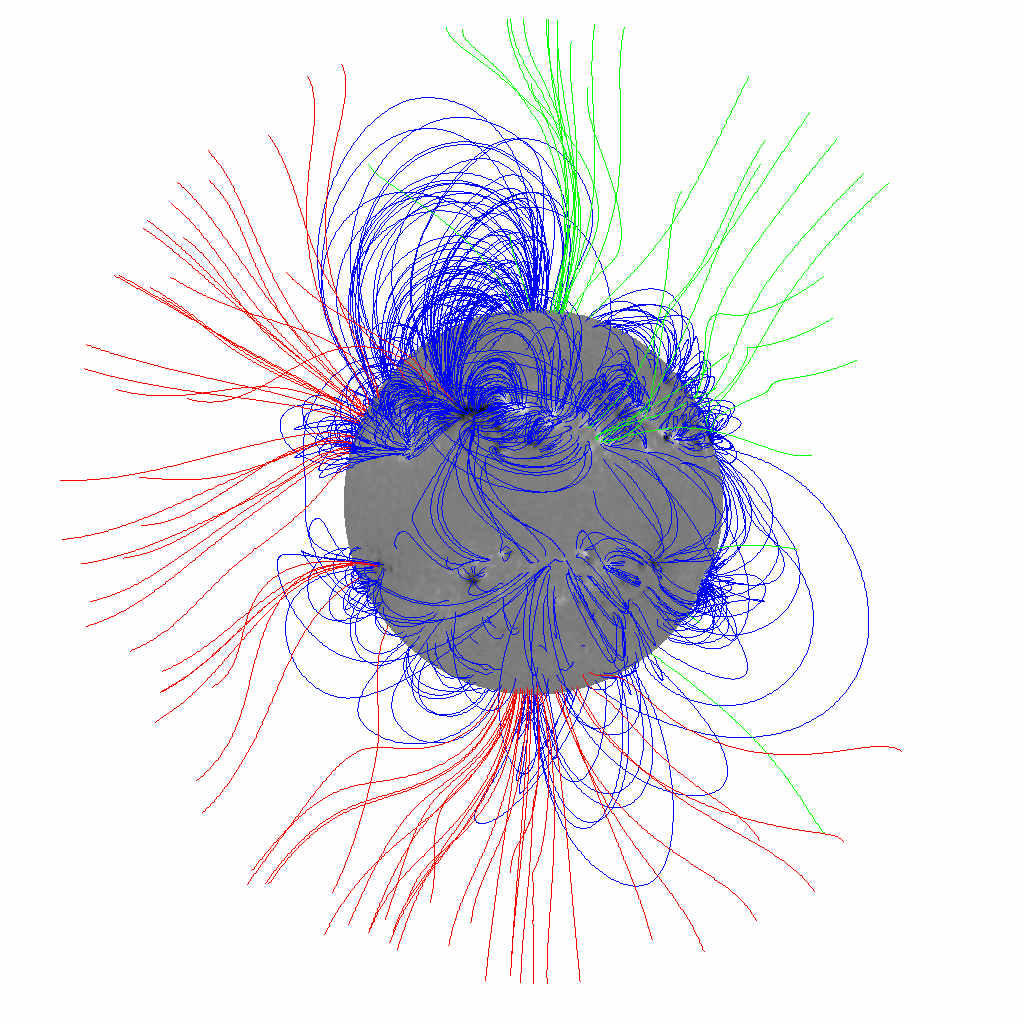}}
\resizebox{0.31\hsize}{!}{\includegraphics*{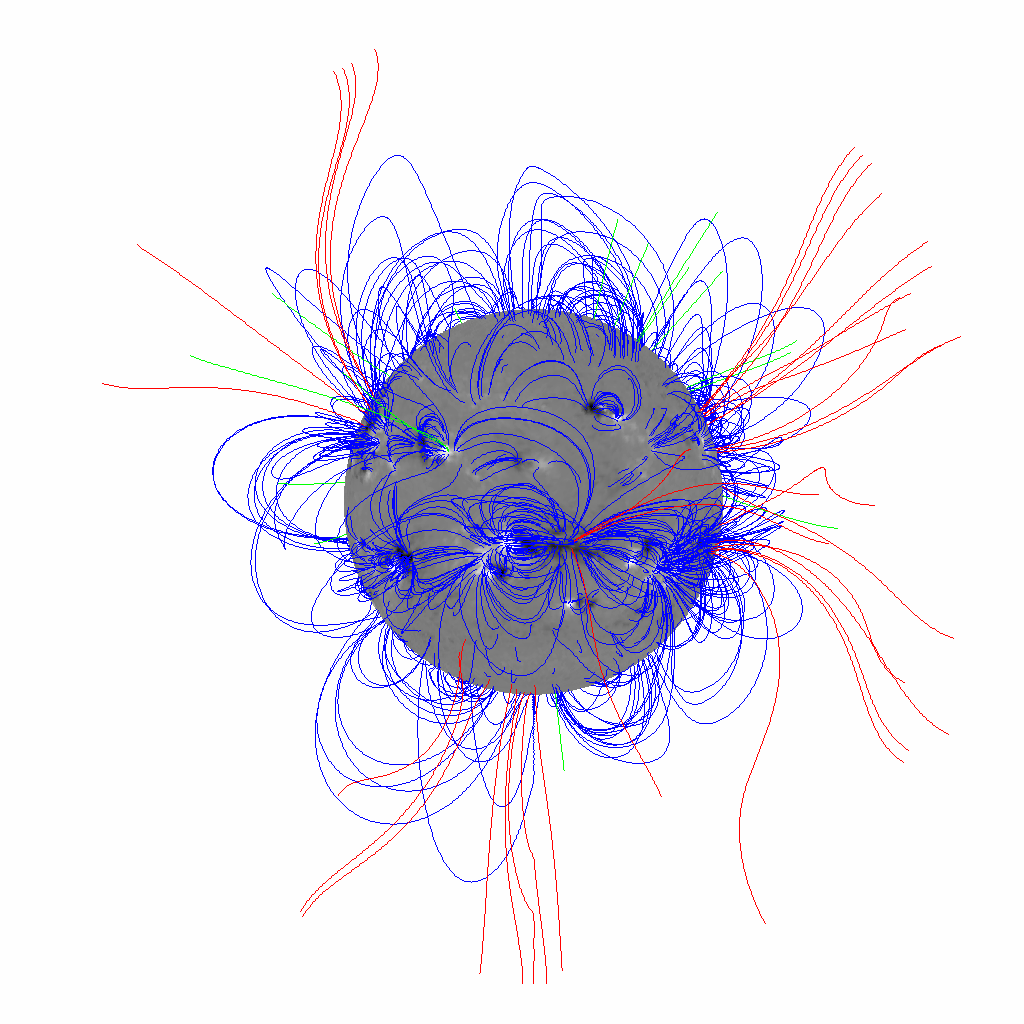}}
\resizebox{0.31\hsize}{!}{\includegraphics*{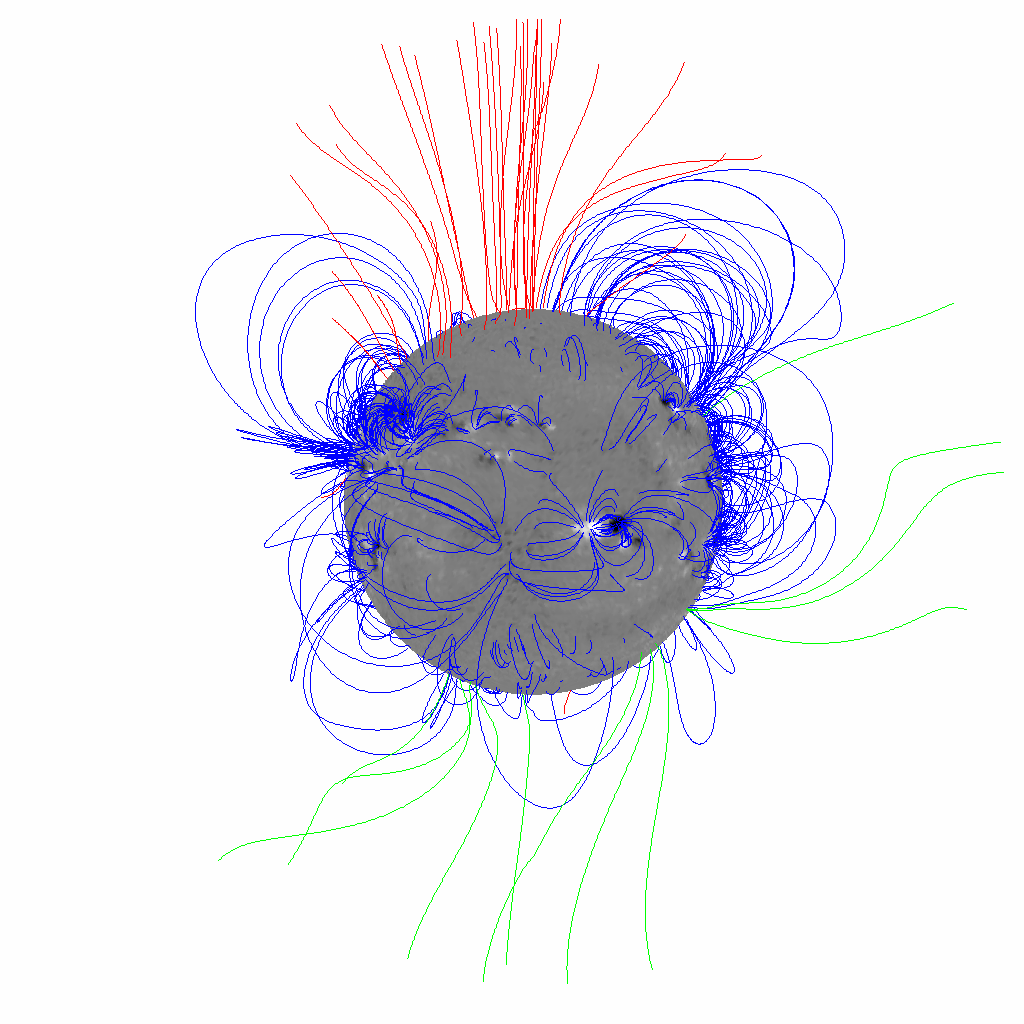}}
\end{center}
\caption{PFSS models representing the beginnings of years 1990-2001. Top row: 1990-92; 2nd row: 1993-95; 3rd row: 1996-98; bottom row: 1999-2001. The photospheric radial field strength is represented by the greyscale, saturated at 100 G with white/black indicating positive/negative polarity. Green/red field lines represent open fields of positive/negative polarity and blue lines represent closed fields.}
\label{fig:pfss1}
\end{figure}

\begin{figure}[h]
\begin{center}
\resizebox{0.31\hsize}{!}{\includegraphics*{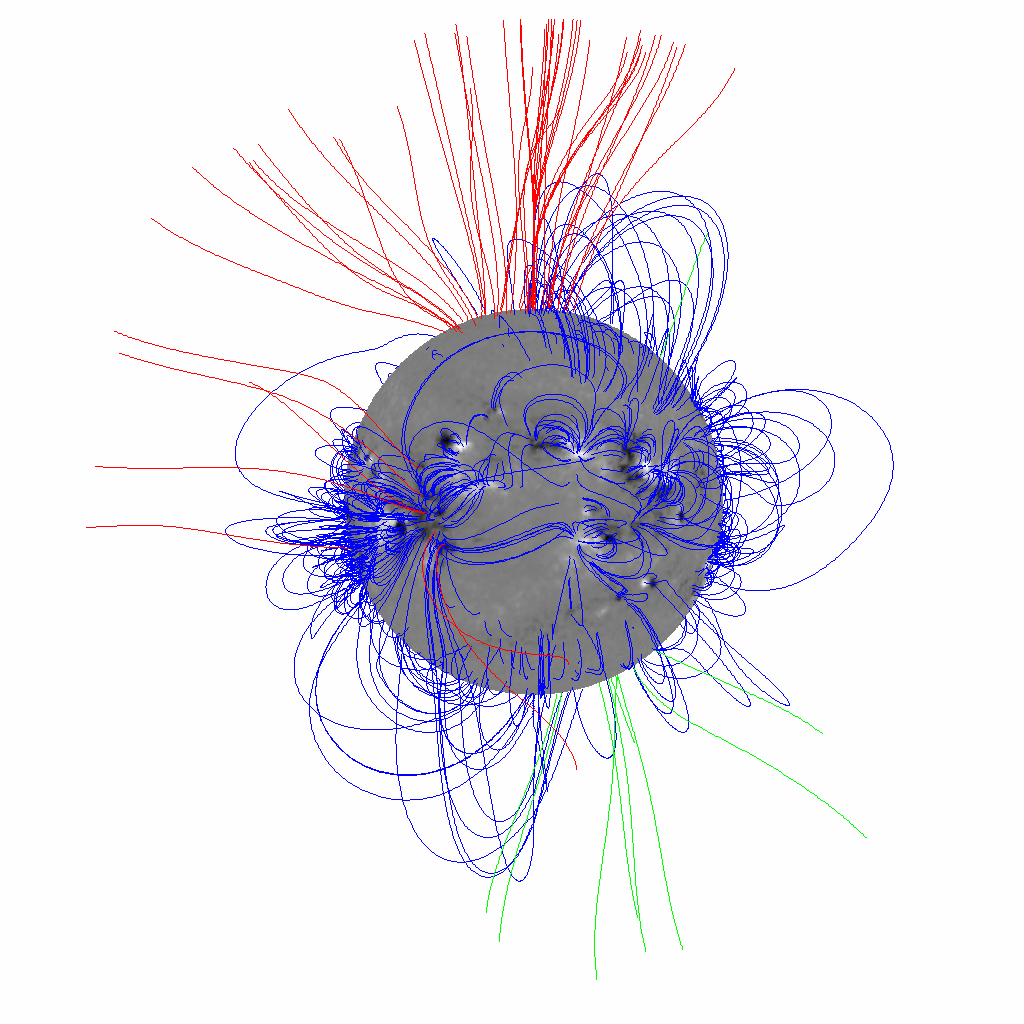}}
\resizebox{0.31\hsize}{!}{\includegraphics*{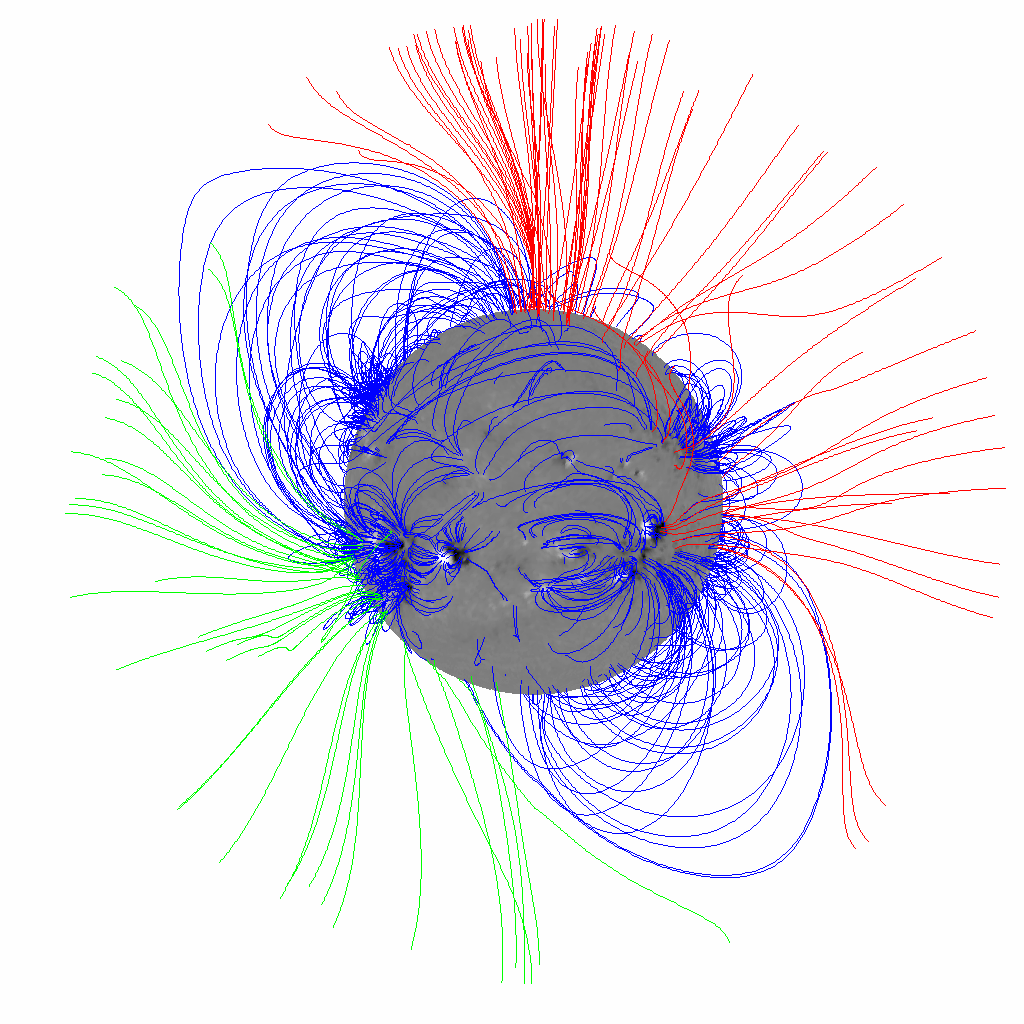}}
\resizebox{0.31\hsize}{!}{\includegraphics*{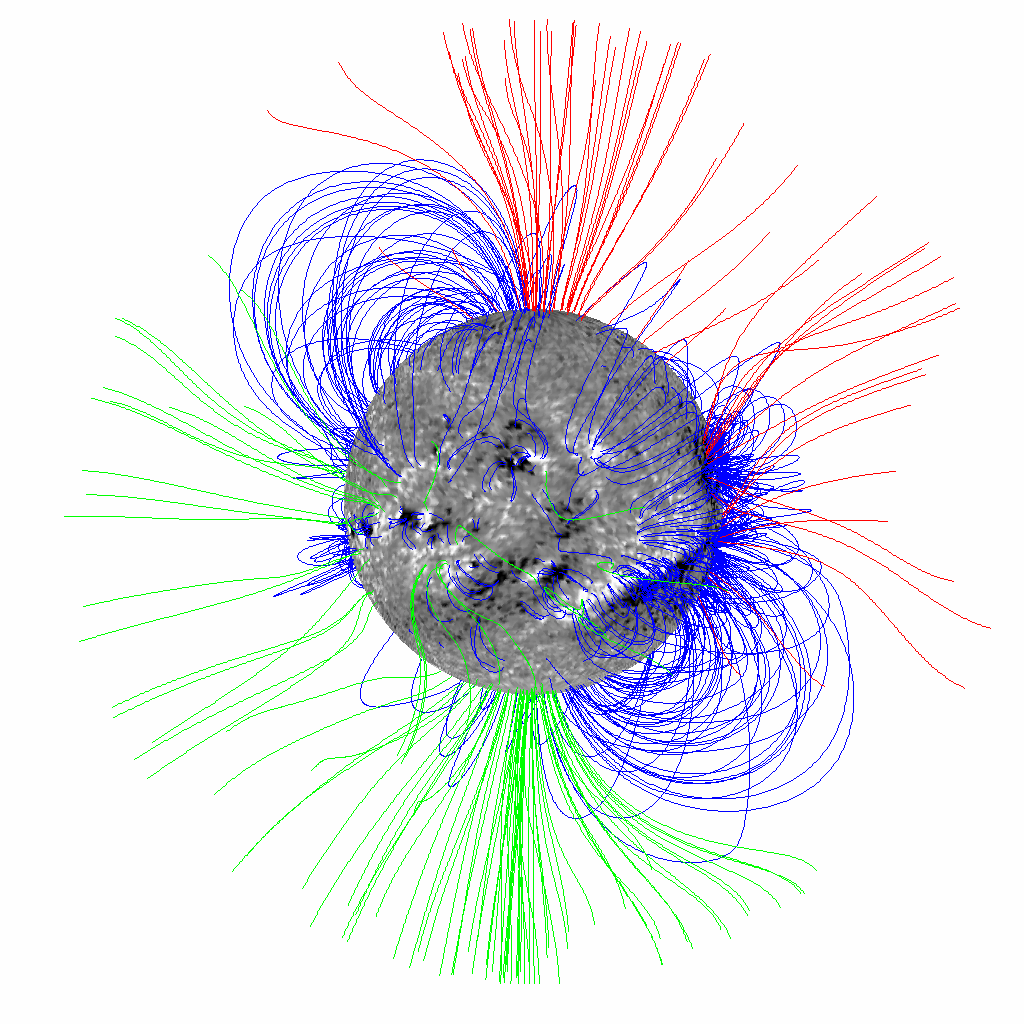}}
\resizebox{0.31\hsize}{!}{\includegraphics*{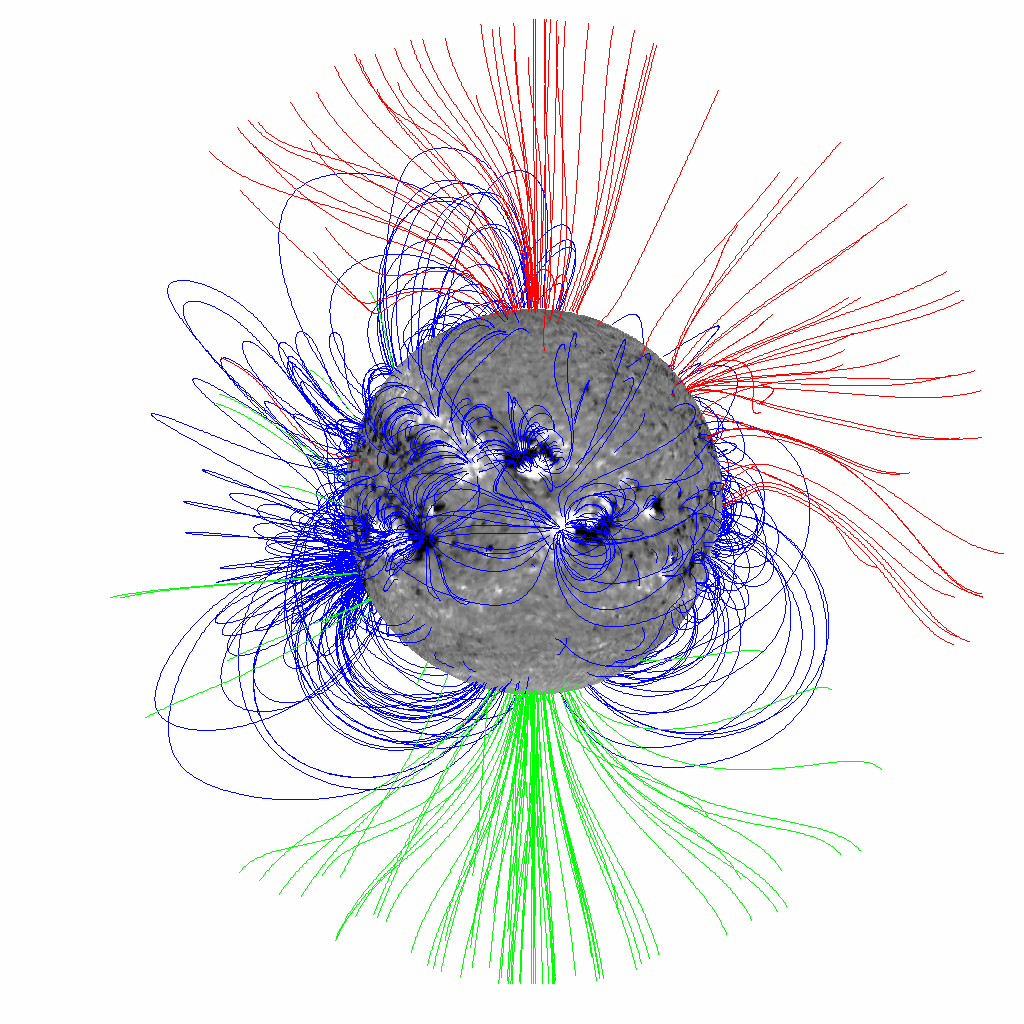}}
\resizebox{0.31\hsize}{!}{\includegraphics*{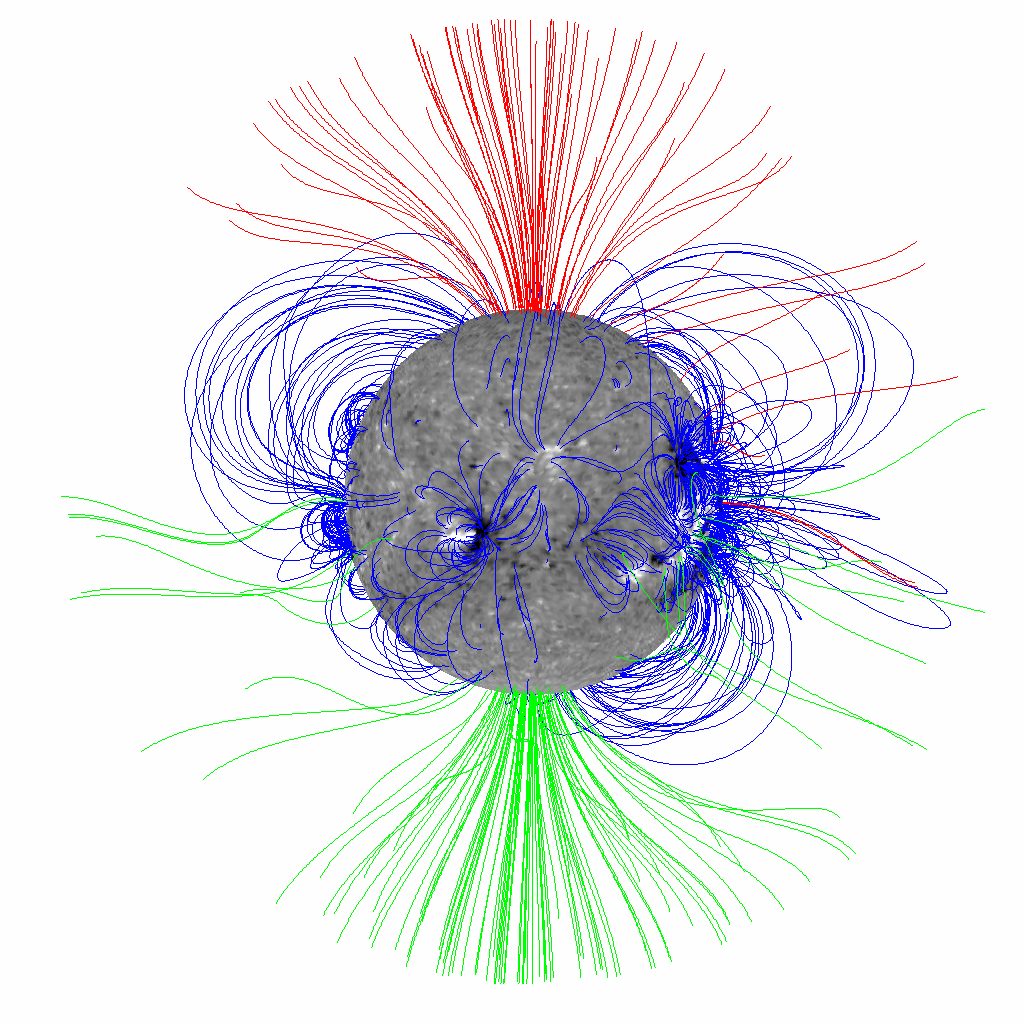}}
\resizebox{0.31\hsize}{!}{\includegraphics*{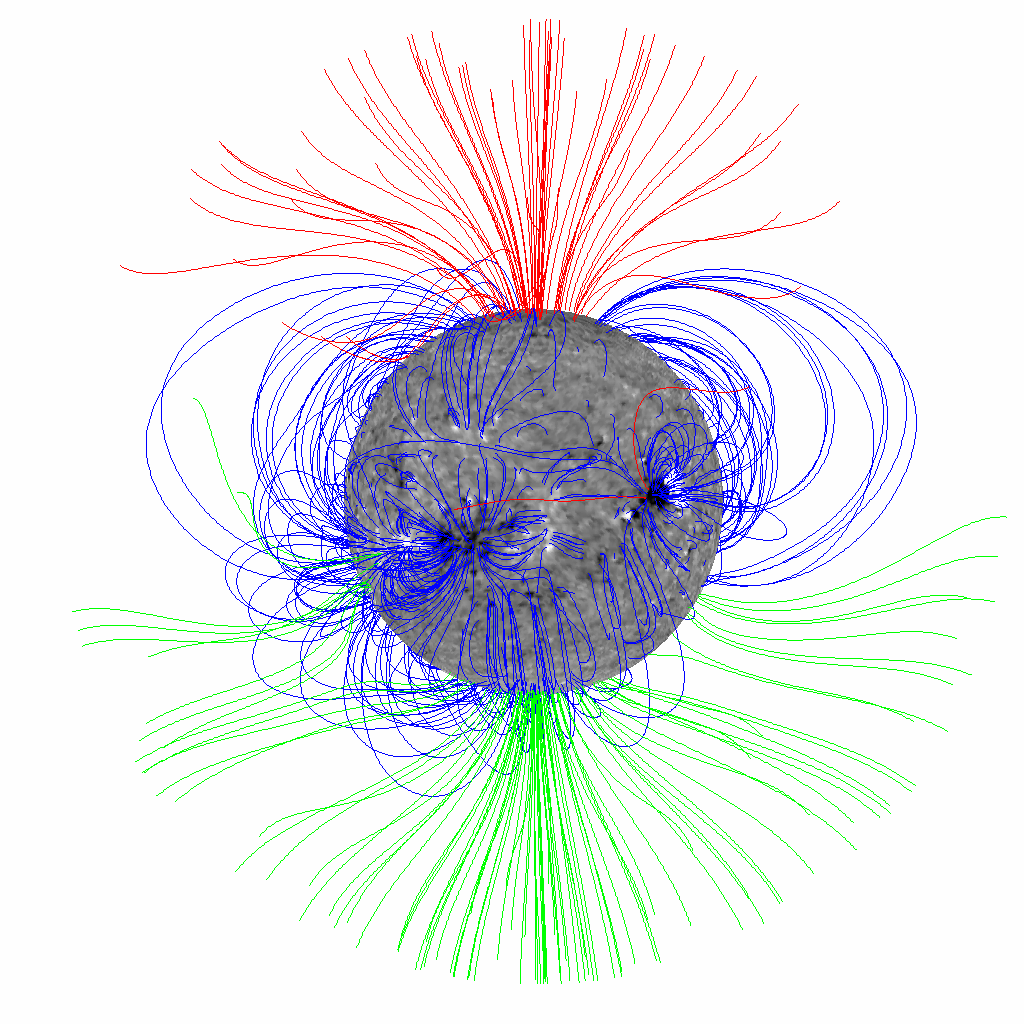}}
\resizebox{0.31\hsize}{!}{\includegraphics*{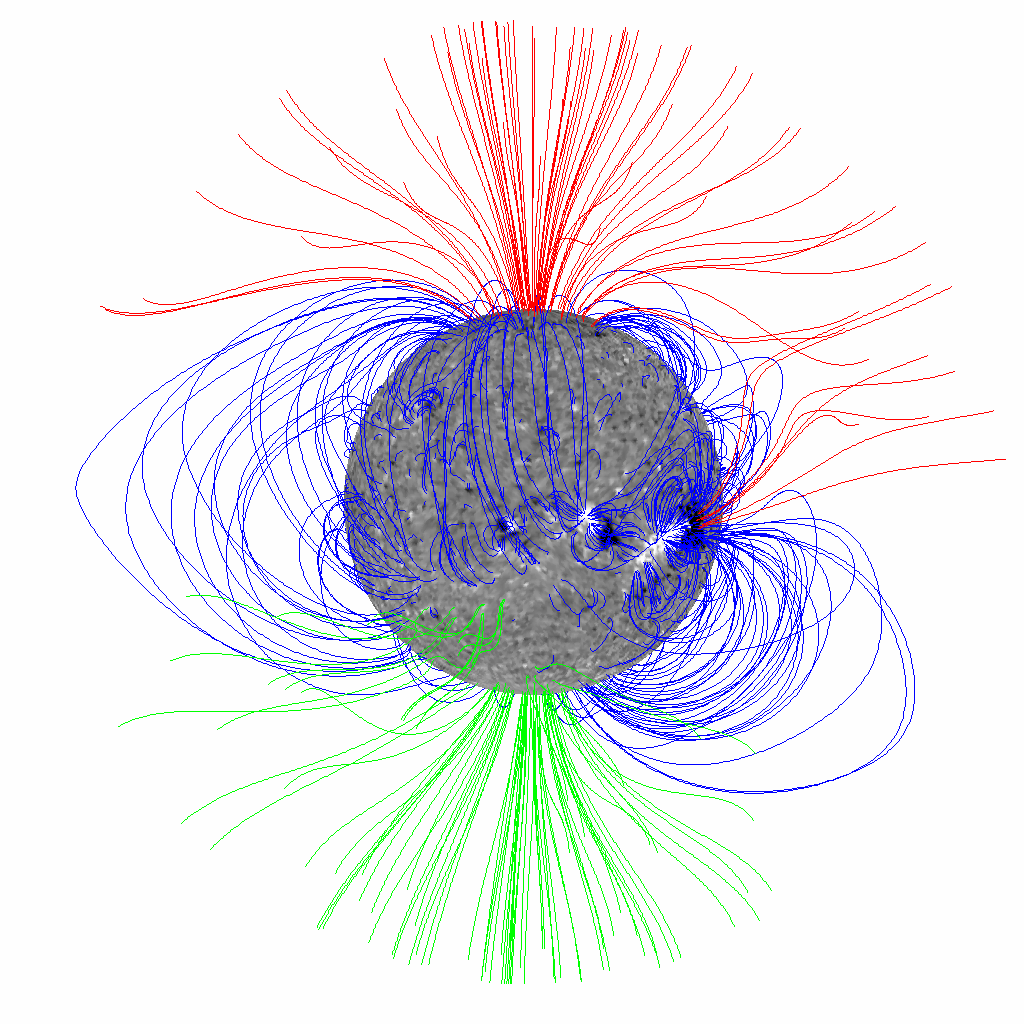}}
\resizebox{0.31\hsize}{!}{\includegraphics*{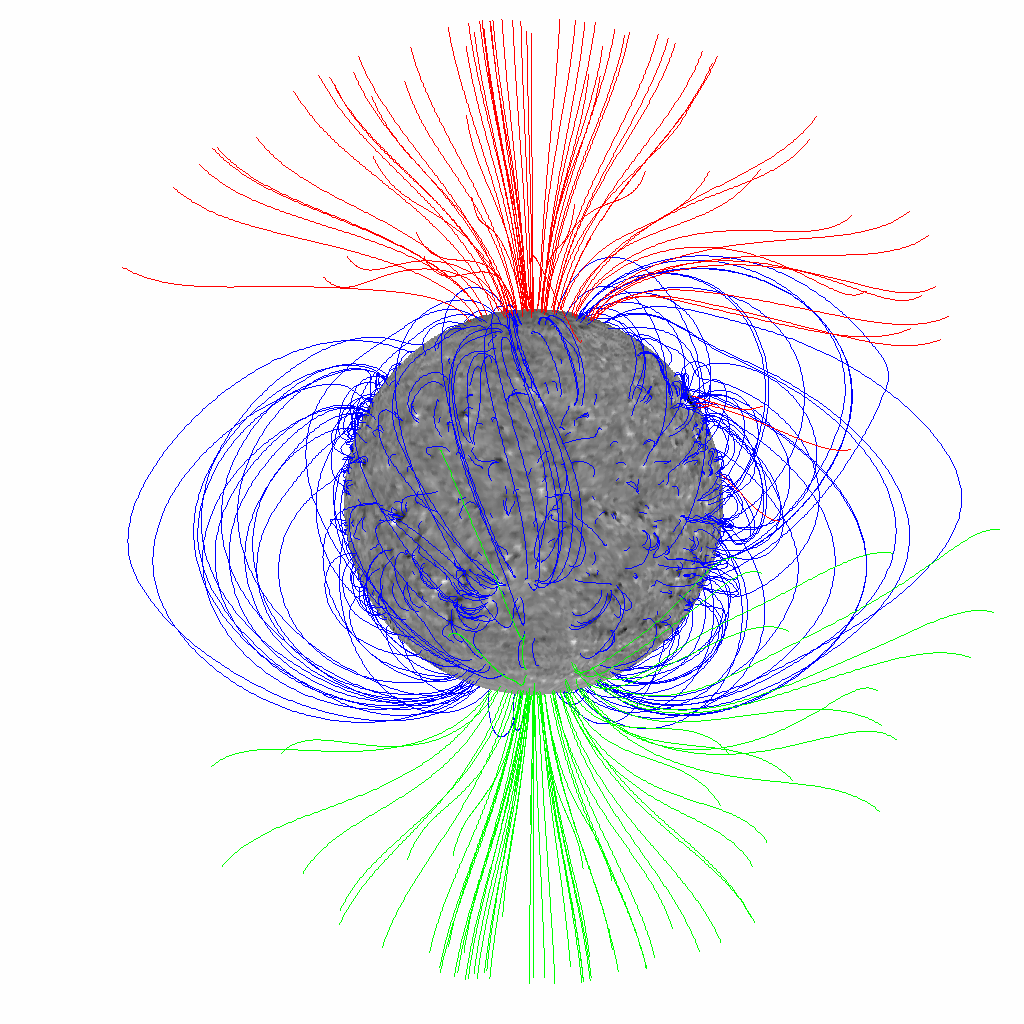}}
\resizebox{0.31\hsize}{!}{\includegraphics*{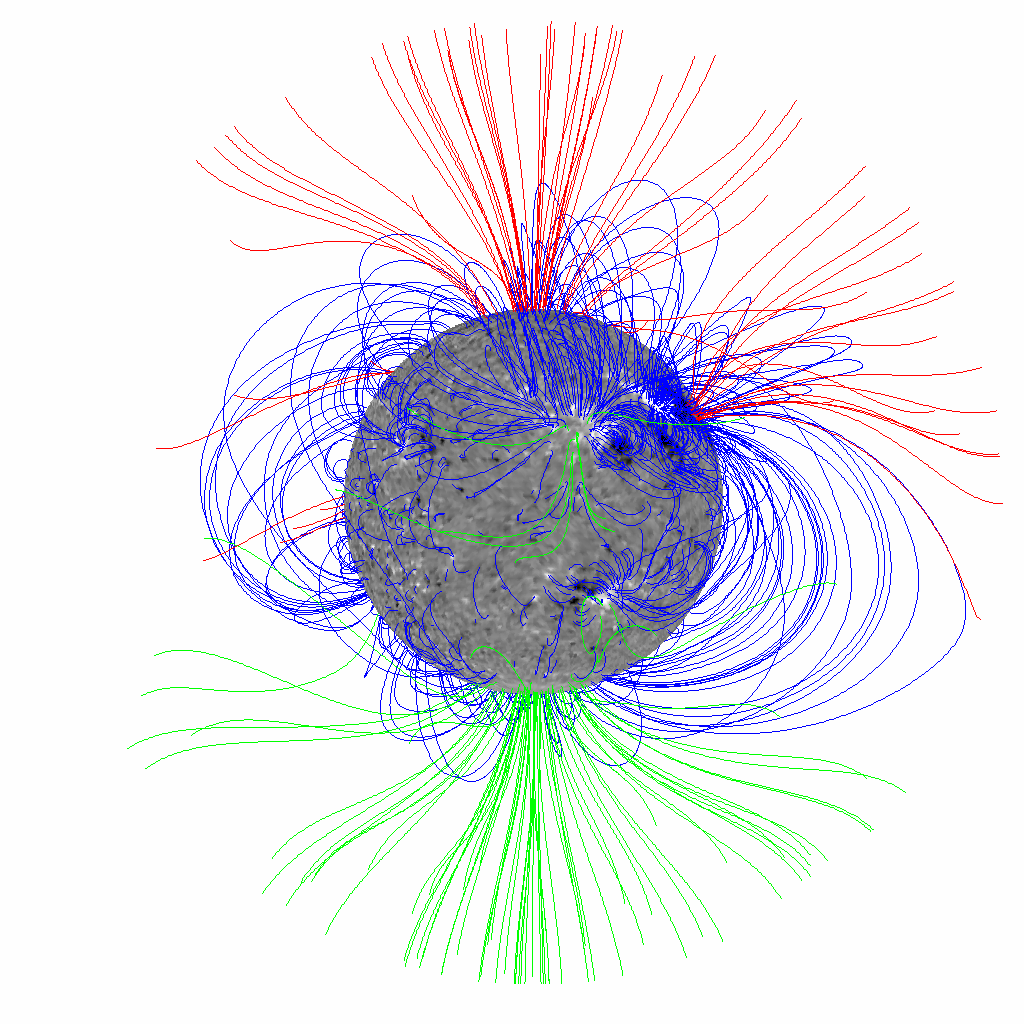}}
\resizebox{0.31\hsize}{!}{\includegraphics*{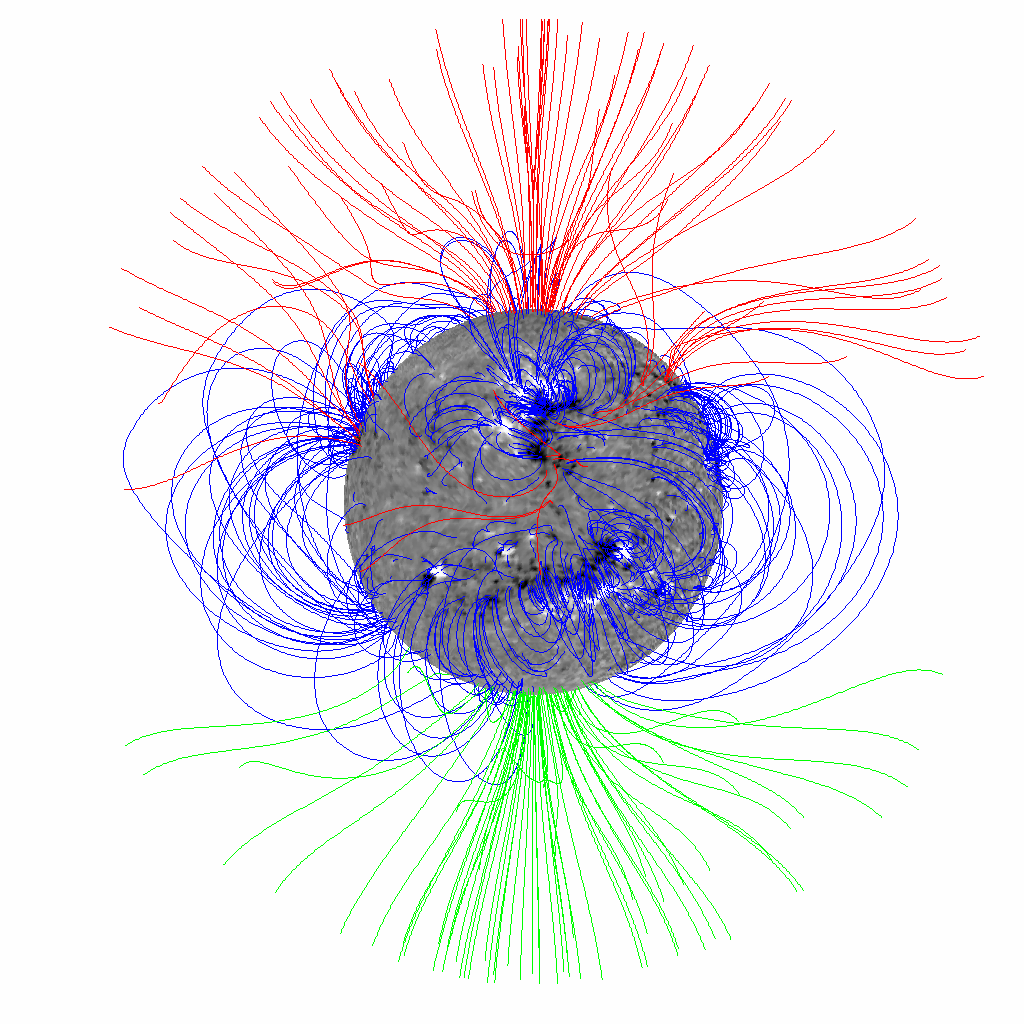}}
\resizebox{0.31\hsize}{!}{\includegraphics*{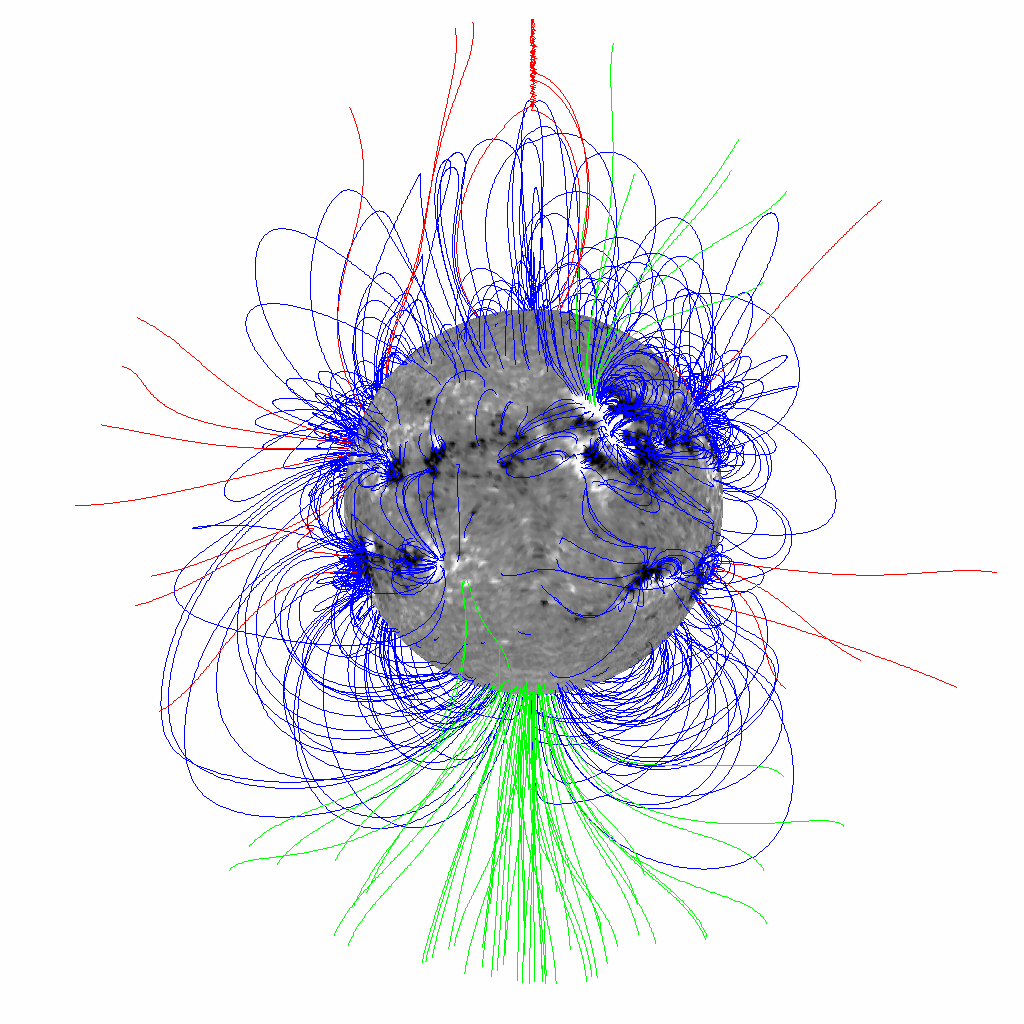}}
\resizebox{0.31\hsize}{!}{\includegraphics*{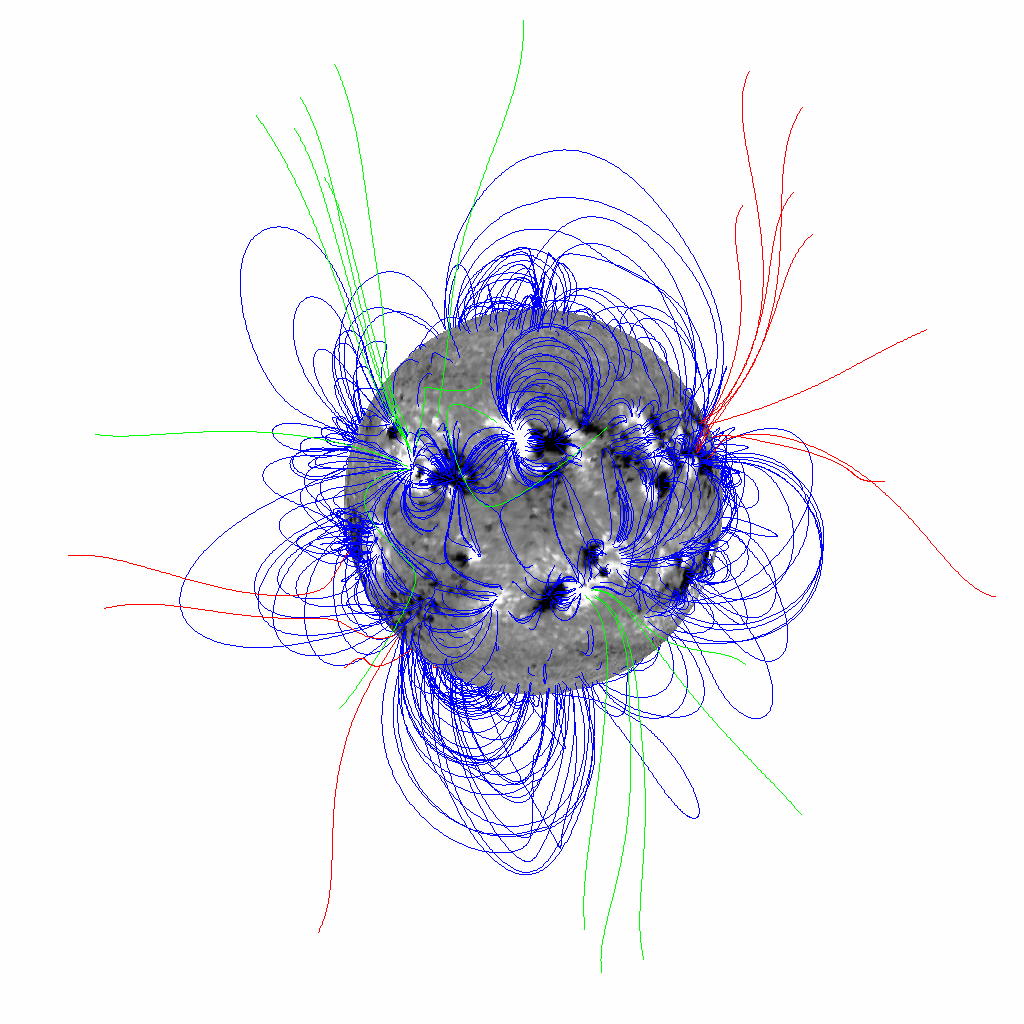}}
\end{center}
\caption{PFSS models representing the beginnings of years 2002-2013. Top row: 2002-04; 2nd row: 2005-07; 3rd row: 2008-10; bottom row: 2011-13. The photospheric radial field strength is represented by the greyscale, saturated at 100 G with white/black indicating positive/negative polarity. Green/red field lines represent open fields of positive/negative polarity and blue lines represent closed fields.}
\label{fig:pfss2}
\end{figure}

Figures~\ref{fig:pfss1} and \ref{fig:pfss2} show hairy-ball plots of 24 example models representing the turns of the 24 years since the beginning of 1990. The foot-points of open fields in these models correspond to coronal holes observed in, e.g., extreme ultraviolet or helium images while the set of tallest closed field lines represent streamer belts seen in coronagraph images. In 1990, corresponding to the beginning of the time series in Figure~\ref{fig:pfss1}, the solar field was close to cycle 22 maximum as Figure~\ref{fig:butterfly} also indicates. In the first model both polar coronal holes are well defined although the northern hole is significantly displaced from the heliographic north pole. In 1991 the polar coronal holes were still present but weaker. By 1992 the polar holes had reversed polarity, and they were already well developed and centered at the heliographic poles. In 1996 activity cycle 22 was over and the coronal field was approximately axisymmetric. By 1998 cycle 23 was well under way and the coronal field was again highly complex and non-axisymmetric. The cycle 23 polar field reversal was different from the cycle 22 reversal. During cycle 23 the north polar coronal hole disappeared about a year earlier than the southern polar hole. This one-year lag can also be seen in Figure~\ref{fig:butterfly}. As Figure~\ref{fig:pfss2} shows, the southern hole was not fully formed until about 2004. Cycle 23 did not produce a nearly axisymmetric, dipole-like global coronal field structure until 2009 when the sunspot number became especially low according to Figure~\ref{fig:ssn}. As has been widely discussed in the literature (see Section~\ref{sect:intro}), this is believed to be due to the weakness of the polar fields during cycle 23. By 2011 cycle 24 had begun and the global coronal field was again complex and non-axisymmetric. The northern polar coronal hole was disappearing by 2012. At the beginning of 2013 the northern polar hole had effectively disappeared, but had not yet reappeared with reversed polarity, while the southern polar hole was weakening steadily.

\subsection{Potential-field source-surface models, spherical harmonics and magnetic multipoles}
\label{s:shpfss}

To investigate the evolution of the coronal field structure in more detail, we also calculate analytical solutions of Equation~(\ref{eq:Laplace}) that are constructed as sums of spherical harmonics. Decomposing global solar magnetic fields into spherical harmonic components reveals which spatial scales and symmetries are dominant during any phase of the field's evolution. These components therefore allow us to summarize the main features of the global field in a useful and instructive way. The WSO group distributes PFSS models and the associated spherical harmonic coefficients on their website\footnote{http://wso.stanford.edu}. We compute spherical harmonic coefficients for the NSO data using the same method, that described by Altschuler and Newkirk~(1969) and Hoeksema~(1984). We briefly describe this method below.

Solutions of Equation~(\ref{eq:Laplace}) in the domain $R\le r\le R_s$ can be written as,

\begin{equation}
\psi (r,\theta ,\phi ) = R\sum_{n=1}^\infty \sum_{m=0}^n \left[ \frac{1}{n+1} \left[{\left(\frac{R}{r}\right)}^{n+1} -{\left(\frac{r}{R_s}\right)}^n {\left(\frac{R}{R_s}\right)}^{n+1} \right] (g_n^m\cos m\phi + h_n^m\sin m\phi ) P_n^m (\theta ) \right] .
\end{equation}

Here the functions $P_n^m (\theta )$ are the usual Legendre polynomials with the Schmidt normalization (Chapman and Bartels~1940, Altschuler and Newkirk~1969, Hoeksema~1984), and the Legendre coefficients $g_n^m$ and $h_n^m$ are in units of magnetic field strength and are determined from the known photospheric field distribution $B_r (R,\theta ,\phi )$ by,

\[\left\{ \begin{array}{c} g_n^m \\ h_n^m \end{array} \right\}  = \frac{2n+1}{4\pi} \int_{\theta=0}^{\pi} \int_{\phi =0}^{2\pi} B_r (R,\theta ,\phi ) P_n^m (\theta )   \left\{ \begin{array}{c} \cos m\phi \\ \sin m\phi \end{array} \right\}  \sin\theta {\mathrm d}\theta {\mathrm d} \phi . \]

With the Schmidt normalization used here for the Legendre polynomials, the coefficients $g_n^m$ and $h_n^m$ refer to the magnitudes of the multipole components of the magnetic field multiplied by $(2n+1)^{1/2}$ (Hoeksema~1984). Because multipoles of different principal index $n$ have different $r$-dependence, the coefficients must also be adjusted by an $r$-dependent factor before multipole strengths can be compared at any chosen height. For example, to measure the comparative influence of two multipoles on the global field structure, their strengths may be compared at $r=R_s$ we must apply the factor $(2n+1)/R_s^{n+2}$ (Hoeksema~1984). The coefficient $g_0^0$ corresponds to the monopole term, set to zero in the computation. The $g_1^0$ is the axial polar dipole (the dipole parallel to the rotation axis) and the terms with coefficients $g_1^1$ and $h_1^1$ represent the two orthogonal equatorial dipoles. The principal index, $n$, is the total number of circles of nodes on the photosphere and the secondary index, $m$, is the number of those nodal circles passing through the pole. Therefore terms with $m=0$ are axisymmetric. The lowest-degree multipoles correspond to the largest length scales, so the comparative amplitudes of the dipole, quadrupole and other low-degree multipoles are useful diagnostics of the global coronal field structure over multiple solar cycles. We describe the temporal evolution of the dipole components in Subsection~\ref{s:dipoles} and the evolution of the  higher-order components in Subsection~\ref{s:multipoles}.





\subsection{Dipolar fields}
\label{s:dipoles}

\begin{figure}[h]
\begin{center}
\resizebox{0.35\hsize}{!}{\includegraphics*{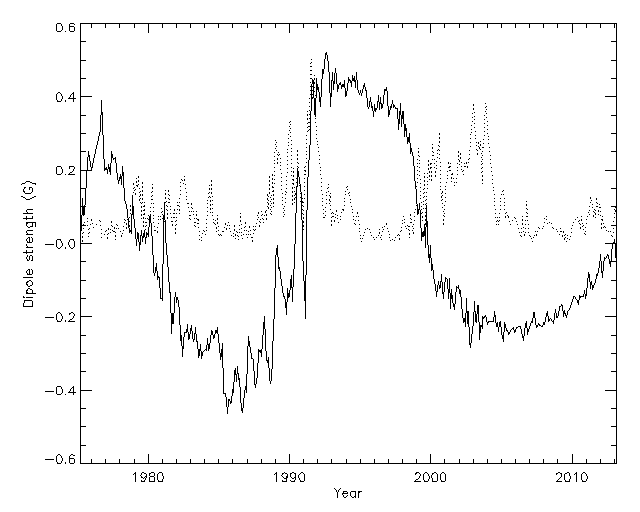}}
\resizebox{0.35\hsize}{!}{\includegraphics*{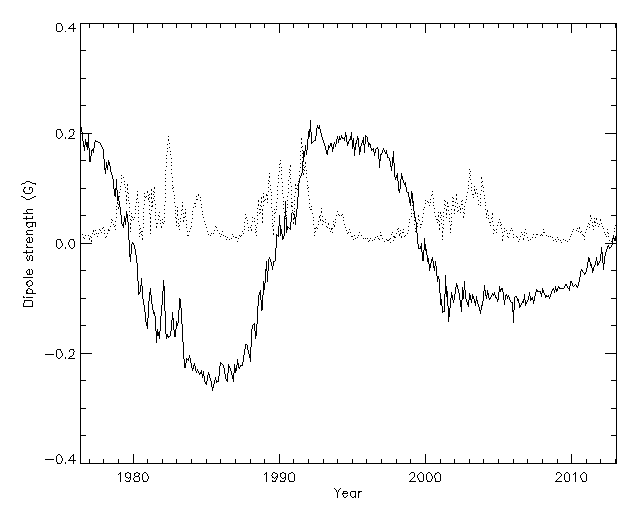}}
\resizebox{0.35\hsize}{!}{\includegraphics*{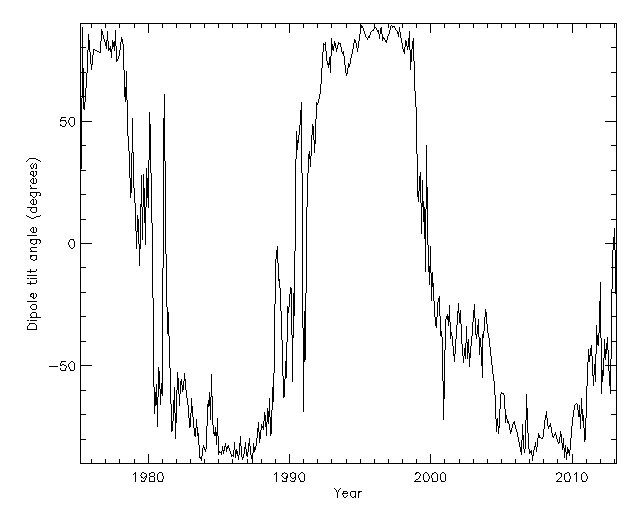}}
\resizebox{0.35\hsize}{!}{\includegraphics*{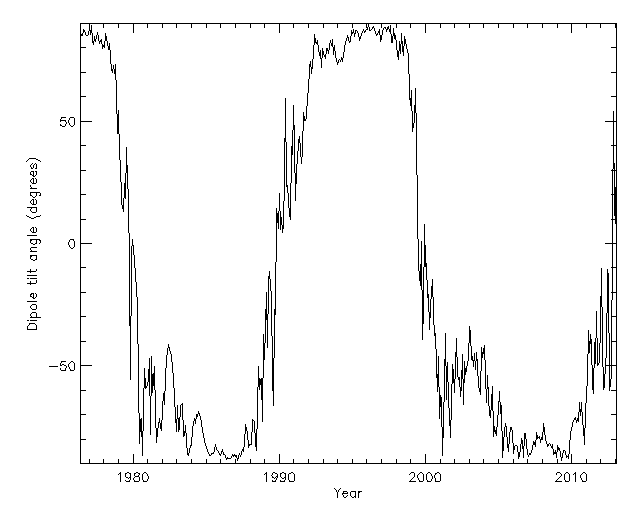}}
\resizebox{0.35\hsize}{!}{\includegraphics*{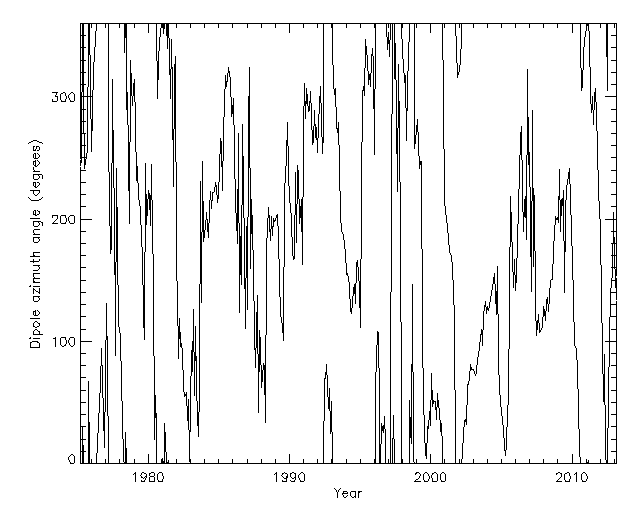}}
\resizebox{0.35\hsize}{!}{\includegraphics*{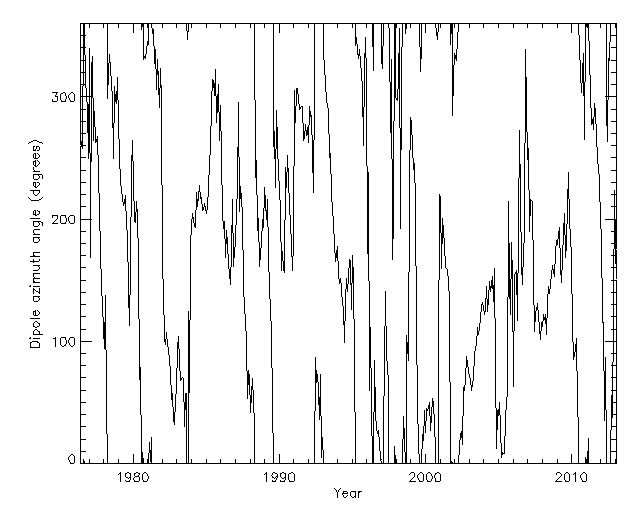}}
\end{center}
\caption{For Kitt Peak (left) and Wilcox (right) data the axial (solid lines) and equatorial (dotted lines) dipole components (top) and the dipole tilt angles (middle) and dipole azimuth angles (bottom) are plotted against time.}
\label{fig:dipoles}
\end{figure}

There are three independent dipole components of the global magnetic field: an axial (axisymmetric) component aligned with the solar rotation axis ($g_1^0$) and two independent components spanning the equatorial plane ($g_1^1$ and $h_1^1$). These dipole components can be added vectorially to give the full strength and direction of the dipole field at any time. The strength of the the equatorial dipole, the vector sum of the equatorial components, can be compared to the axial dipole as shown in Figure~\ref{fig:dipoles}. In this figure the axial dipole points north when positive and south when negative. The axial dipole is observed to be at its strongest during solar minimum, after the previous cycle's decayed field has accumulated at the poles but before the new cycle has begun to send opposite-polarity field poleward weakening the polar fields towards polarity reversal. The equatorial dipole, on the other hand, follows the activity cycle, being largest during activity maximum and smallest during activity minimum. Figure~\ref{fig:dipoles} shows that the axial dipole has been becoming progressively weaker from cycle to cycle since cycle 21. This evolution follows the pattern of the polar field strengths plotted in Figure~\ref{fig:poles}. The equatorial dipole has also been weaker during cycles 23 and 24 than during cycles 21 and 22, following the behavior of the sunspot number shown in Figure~\ref{fig:ssn}. The equatorial dipole is formed mostly by the azimuthal (east-west) component of active region fields, whenever these form a net dipole moment in the equatorial plane. Because of the Joy's law tilt bias in each hemisphere, the active regions would tend to produce a small axial dipole moment of the same sign in each hemisphere but this moment is small, and is insignificant compared to the moment of the polar fields because of the generally small Joy's law tilt angle. The influence of the active regions on the equatorial dipole is much larger. The effect of the polar fields on the equatorial dipole, through non-axisymmetry in the polar fields, is insignificant as Figure~\ref{fig:dipoles} indicates.

The dipole tilt angle, the angle in the poloidal plane between the dipole and the equator, is also plotted in Figure~\ref{fig:dipoles}. This is a useful diagnostic of the global coronal field because it offers a comparison of the polar and low-latitude (active region) contributions to the global field. The tilt is close to $\pm 90^{\circ}$ latitude during solar minimum and passes through $0^{\circ}$ at the height of maximum activity when the polar fields reverse polarity. This process can take a few months to a year at each pole, and the two poles do not always reverse polarity around the same time. The axial dipole reversals have taken between one and two years with the cycle 22 reversal around 1990 the quickest and simplest of the sample, corresponding to the comparatively quick and simple cycle 22 reversal seen in Figures~\ref{fig:butterfly} and \ref{fig:pfss1}. The dipole strength is at its weakest during polar reversal but it does not disappear at any time.

Figure~\ref{fig:dipoles} also shows the azimuthal angle of the horizontal dipole. This angle indicates which Carrington longitude the horizontal dipole is pointing towards. For example, the dipole azimuthal angle is zero when the dipole points towards $0^{\circ}$ Carrington longitude, the boundary between two Carrington rotations. Figure~\ref{fig:dipoles} shows that the dipole azimuth angle has no significant dependence on the activity cycle. The angle often changes abruptly, on time-scales much less than a year, as different active regions emerge and decay at different Carrington longitudes. There are, however, periods of time such as during 1984, 1991, 2008 and 2009, for example, when the dipole maintains a nearly stable azimuthal direction over timescales of about a year. This is because during these periods of time most of the activity appears preferentially within a small range of Carrington longitudes. There are numerous episodes evident in Figure~\ref{fig:dipoles} when the activity appears to have been dominated by such ``active longitudes'' but such behavior shows no dependence on the activity cycle.

One remarkable feature of Figure~\ref{fig:dipoles} is that the active field, represented by the equatorial dipole component, continued to appear with significant strength during the declining phase of cycle 23 long after the polar fields ceased to strengthen. The polar dipole component  became stronger between the reversal of the poles in 1998 and 2001 but after 2002 the polar dipole did not become stronger. On the other hand, the active fields continued to appear long after 2002, until 2006, without having any significant net effect on the polar dipole. This pattern can also be seen in the dipole tilt angle. After 2002, after the poles reversed and the dipole tilt swung decisively southward, the trend of the tilt angle halted and even reversed for a few years. During this time the polar dipole did not strengthen while the equatorial dipole, representing the active fields, continued to be strong. Indeed, it was arguably stronger in 2003 and 2004, after the polar dipole stopped developing, than before, as Figure~\ref{fig:dipoles} shows.  Between about 2001 and 2004, after the cycle 23 polar field reversal, the equatorial dipole remained comparable in strength to the polar dipole. An essential part of the Babcock-Leighton mechanism is the change in the polar field strength caused by the poleward drift of decayed active-region flux. Between 2002 and 2006 active-region flux continued to appear in large quantities but the polar fields did not change accordingly. We will return to this issue in the following subsection.


More recently the polar dipole has steadily weakened as the Cycle 24 polar field reversal approaches. Meanwhile the equatorial dipole has increased, reflecting the ascent of cycle 24.  The dipole tilt angle is quickly decreasing, principally because of the weakening polar dipole. According to the latest synoptic data from both SOLIS and Wilcox the dipole tilt has recently turned positive and the polar dipole has recently pointed north for the first time in this cycle.


\clearpage

\subsection{Multipolar fields}
\label{s:multipoles}

\begin{figure}[h]
\begin{center}
\resizebox{0.35\hsize}{!}{\includegraphics*{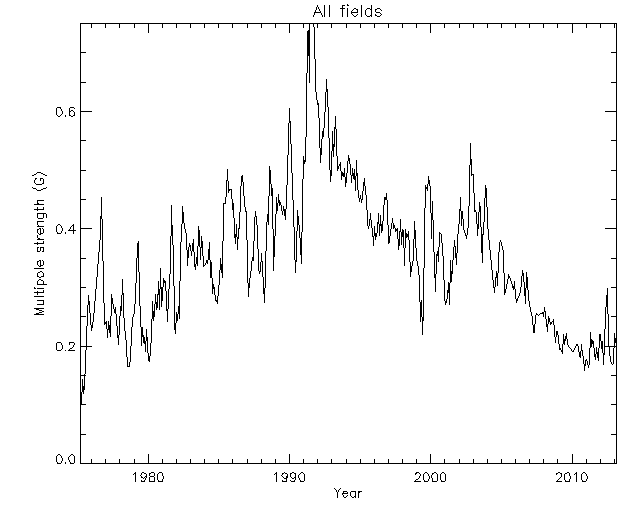}}
\resizebox{0.35\hsize}{!}{\includegraphics*{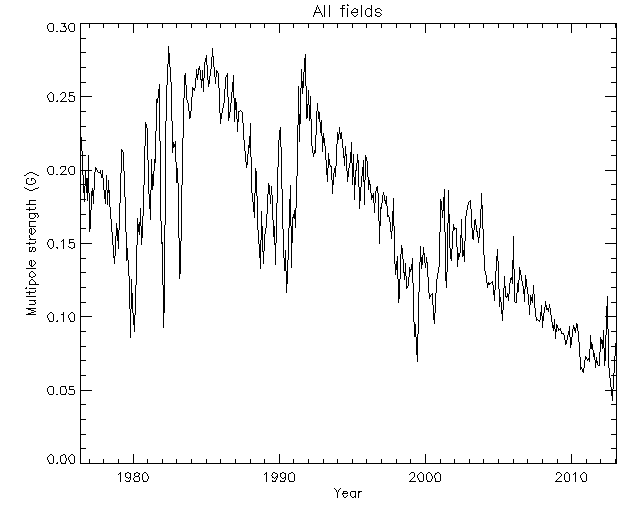}}
\resizebox{0.35\hsize}{!}{\includegraphics*{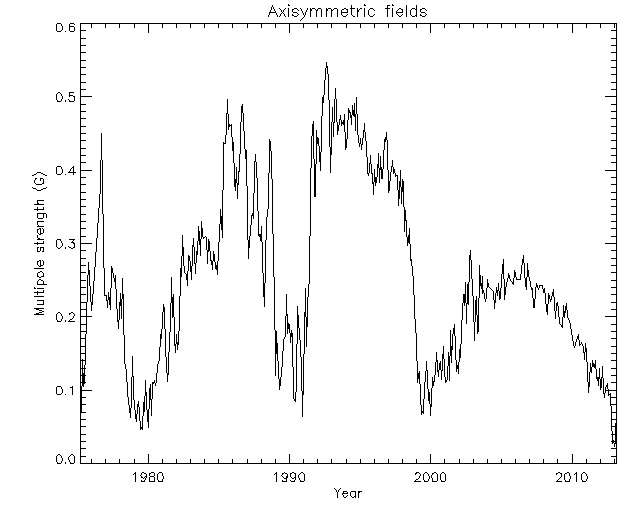}}
\resizebox{0.35\hsize}{!}{\includegraphics*{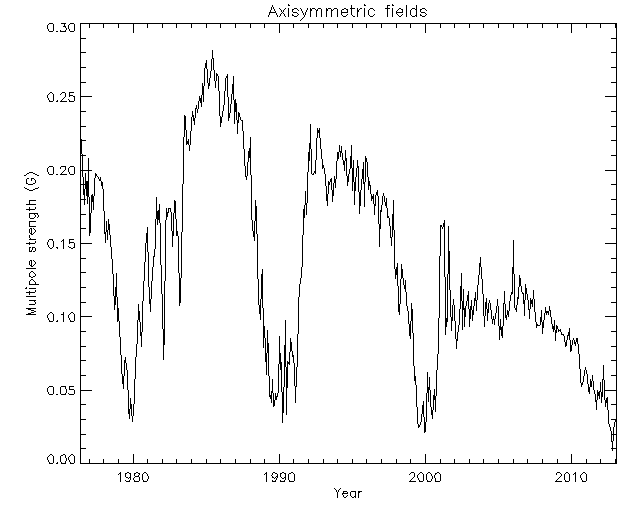}}
\resizebox{0.35\hsize}{!}{\includegraphics*{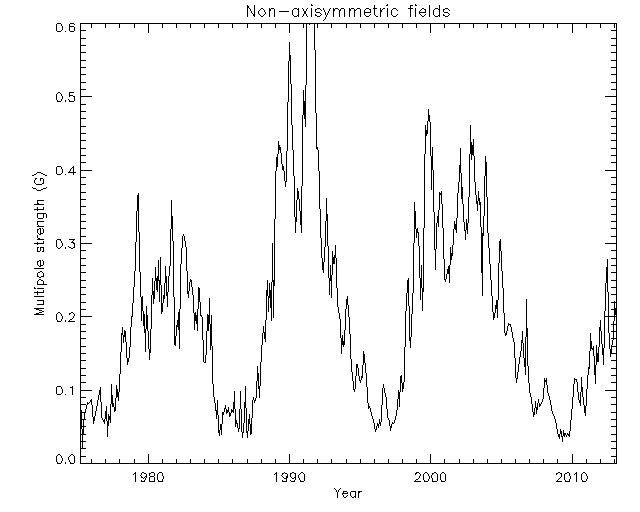}}
\resizebox{0.35\hsize}{!}{\includegraphics*{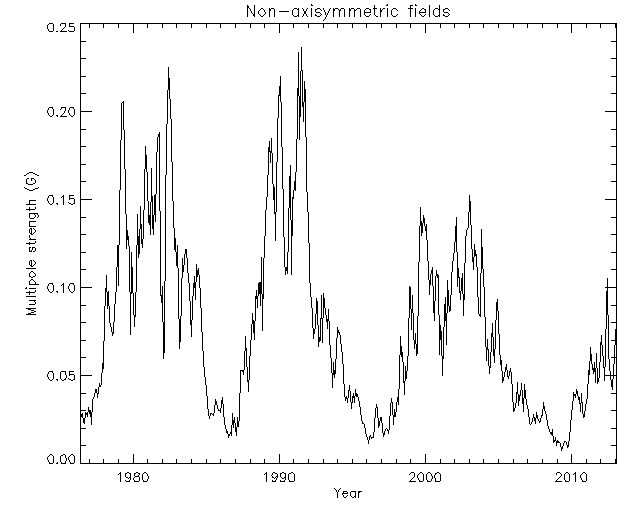}}
\end{center}
\caption{For Kitt Peak (left) and Wilcox (right) data the total field strengths (top), the axisymmetric field strengths (middle) and the non-axisymmetric field strengths (bottom) at 2.5 solar radii are plotted against time.}
\label{fig:mpoles}
\end{figure}


\begin{figure}[h]
\begin{center}
\resizebox{0.95\hsize}{!}{\includegraphics*{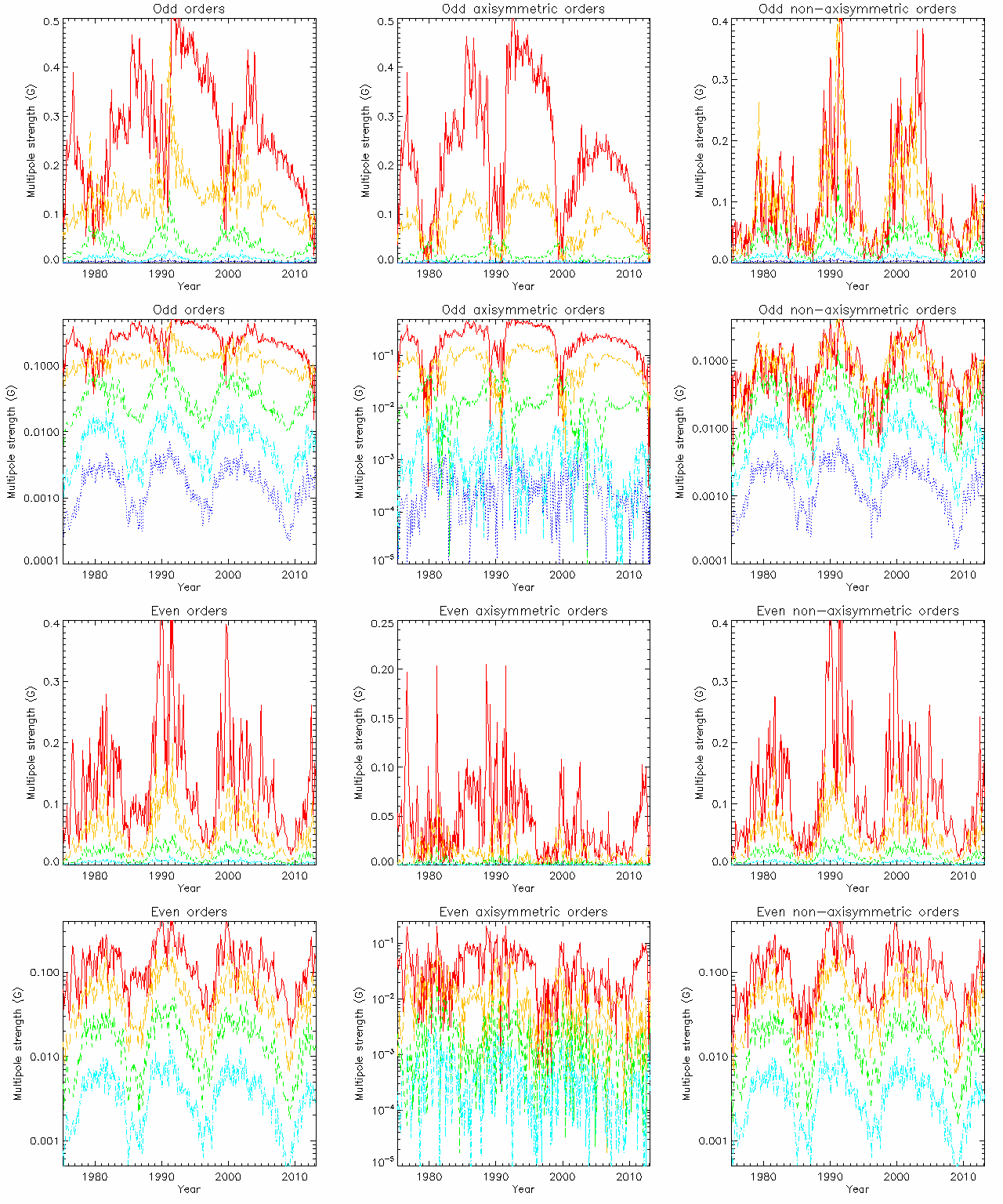}}
\end{center}
\caption{For Kitt Peak data the first several even- and odd-order multipole components are plotted separately, including all, axisymmetric or non-axisymmetric fields as indicated. For odd orders, red, amber, green, cyan and blue represent fields with $n=1, 3, 5, 7, 9$, respectively. For even orders, red, amber, green and cyan represent fields with $n=2, 4, 6, 8$, respectively.}
\label{fig:mpolesarraykp}
\end{figure}


\begin{figure}[h]
\begin{center}
\resizebox{0.95\hsize}{!}{\includegraphics*{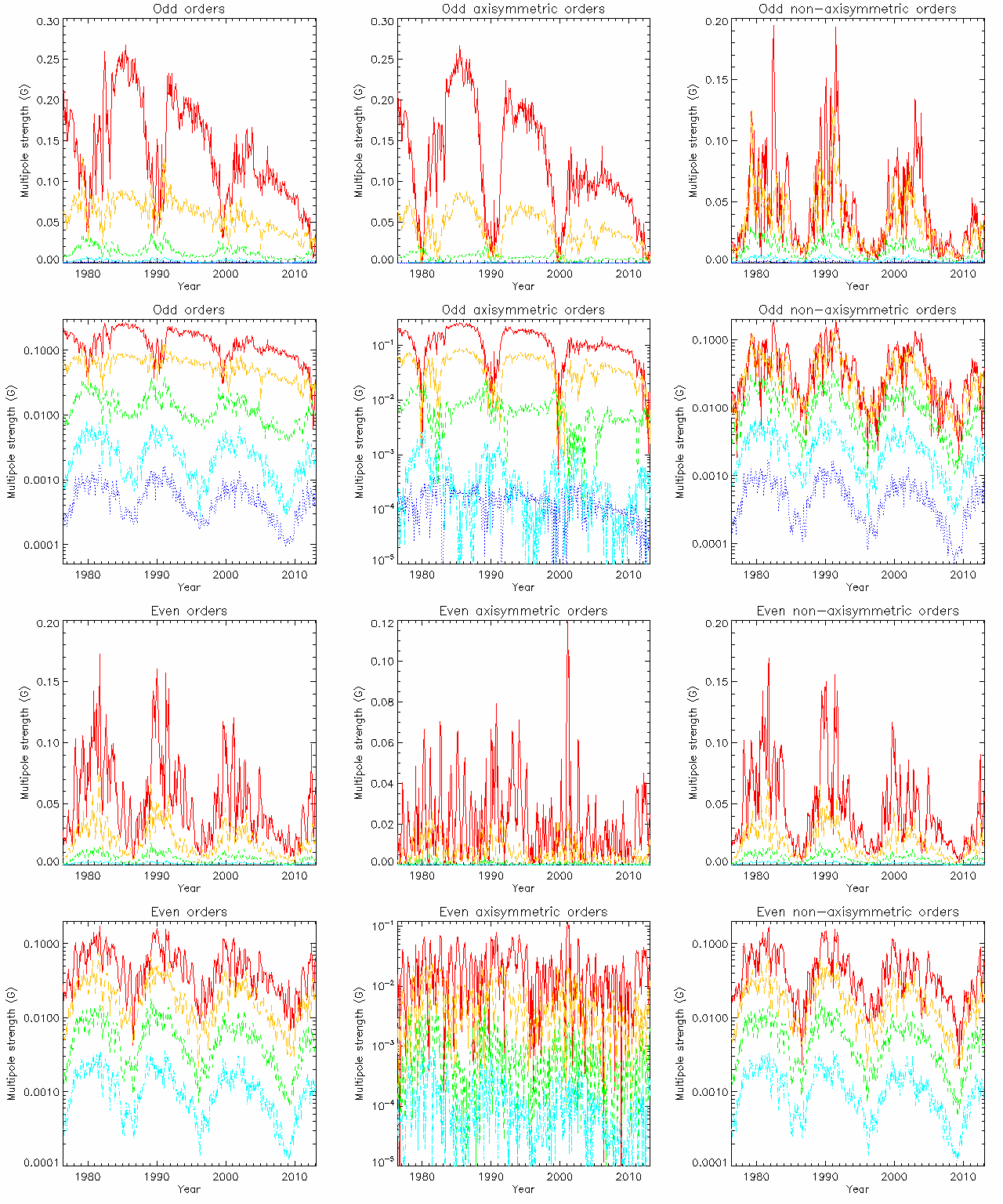}}
\end{center}
\caption{For Wilcox data the first several even- and odd-order multipole components are plotted separately, including all, axisymmetric or non-axisymmetric fields as indicated.  For odd orders, red, amber, green, cyan and blue represent fields with $n=1, 3, 5, 7, 9$, respectively. For even orders, red, amber, green and cyan represent fields with $n=2, 4, 6, 8$, respectively.}
\label{fig:mpolesarraywilcox}
\end{figure}

The simplest way to understand the cyclic behavior of the higher-order multipoles is to plot the axisymmetric ($m=0$) and non-axisymmetric ($m\ne 0$) fields separately. Figure~\ref{fig:mpoles} shows plots of the full field strength, the combined strength of all multipoles, as well as the strength of the axisymmetric and non-axisymmetric components, measured by the two observatories. Figure~\ref{fig:mpoles} shows that the axisymmetric fields dominate during activity minima and the non-axisymmetric fields dominate during maxima. The Pearson linear correlation coefficient ($cc$) between the time series for all WSO fields and the sunspot number (Figure~\ref{fig:ssn}) is low, about 0.10. The $cc$ between all WSO fields and the polar field strength, the combined strengths of the north and south polar fields plotted in Figure~\ref{fig:poles}, is about 0.75. The $cc$ between the non-axisymmetric WSO fields and the sunspot number is much higher, about 0.97. The axisymmetric WSO fields correlate equally well with the total polar field strength with $cc$ about 0.97. The axisymmetric fields began to decline slowly after 2007. Figure~\ref{fig:poles} shows that this may be because the north polar field began to weaken steadily after 2007. The correlations for the NSO/KP data are less impressive because of the difficulty of cross-calibration between the three NSO/KP magnetographs (see Section~\ref{sect:data}) but the patterns of correlation are qualitatively the same. These correlations illustrate the enormous influence of the polar fields on the global coronal magnetic field, and suggests an association between sunspot/active region fields and non-axisymmetric global coronal structure on the one hand, and polar fields and axisymmetric global coronal structure on the other. This association is not simple, however, as the following discussion will show.

In Figures~\ref{fig:mpolesarraykp} and \ref{fig:mpolesarraywilcox} the individual poloidal orders are plotted separately. Higher-order multipoles differ in behavior from the dipole ($n=1$) components in general. For example, the full quadrupolar fields ($n=2$, Figures~\ref{fig:mpolesarraykp} and \ref{fig:mpolesarraywilcox}, top left plots, red curves) differ from the dipolar fields ($n=1$, Figures~\ref{fig:mpolesarraykp} and \ref{fig:mpolesarraywilcox}, bottom left plots, red curves) in being strongest during activity maxima and weakest during activity minima. The axisymmetric and non-axisymmetric quadrupolar fields follow the activity cycle in this way. Unlike the axial dipole (top middle plots, red curves), the axisymmetric quadrupole (bottom middle plots, red curves) is of even order and is symmetric about the equator, and therefore its behavior is not dominated by the polar fields. All quadrupolar components, axisymmetric and non-axisymmetric, are perturbed mostly by the active region fields and their strengths wax and wane with the activity cycle. The octupole fields are of odd order ($n=3$) and so the axisymmetric octupole, like the axial dipole, follows the behavior of the polar fields while the non-axisymmetric octupolar fields follow the activity cycle.

The qualitative patterns described here apply to both NSO/KP and WSO data sets but the correlation coefficients quoted below derive from the WSO data only. The high-order fields are generally well correlated with the sunspot number. The full even- and odd-order multipoles correlate equally well ($cc\approx 0.96$) with the sunspot number for orders higher than hexadecapole ($n=4$). The low-order even multipoles also correlate well ($cc > 0.9$) with the sunspot number whereas the low-order odd multipoles, the dipole ($n=1$) and octupole ($n=3$), instead follow the polar fields. The axisymmetric dipole ($cc=0.97$) and octupole ($cc=0.96$) components correlate well with the polar fields but the axisymmetric multipoles of higher order do not ($cc<0.1$). The low even axisymmetric orders are not correlated with the sunspot number or with the polar fields.  The high even axisymmetric orders show some correlation with the sunspot number ($cc\approx 0.8$ in some cases), possibly perturbed by Joy's law tilt of active region fields, but these components are weaker than their non-axisymmetric counterparts. Therefore the excellent overall correlation between the axisymmetric fields and the polar fields is exclusively due to the axial dipole and octupole fields. Because of their large spatial scale and impressive strength these fields have a major influence on the global coronal field over most of the cycle.

The non-axisymmetric multipole fields correlate increasingly well with the sunspot number as the order $n$ increases, from $cc=0.82$ for the equatorial dipole to $cc=0.97$ for the ninth-order fields. This behavior corresponds to the patterns in the right plots in Figures~\ref{fig:mpolesarraykp} and \ref{fig:mpolesarraywilcox}, where in particular the response of the equatorial dipole to the activity cycle is clearly not as steady as the response of the higher-order non-axisymmetric multipoles. Since low multipole orders correspond to large spatial scales, it may frequently occur that multiple active regions perturb high-order multipoles without the lowest-order multipoles responding significantly. For example, two identical active regions at antipodal points would give zero equatorial dipole but non-zero higher multipole components. There is no evidence that the non-axisymmetric multipoles respond to the polar field cycle, including during polar reversal. Figures~\ref{fig:pfss1} and \ref{fig:pfss2} clearly show that the lowest-order dipole-like structure of the global coronal field is often tilted with respect to the rotation axis. The centers of the polar coronal holes are sometimes significantly displaced from the heliographic poles. The absence of any significant polar field signature in the non-axisymmetric fields in Figures~\ref{fig:mpolesarraykp} and \ref{fig:mpolesarraywilcox} shows that the polar field reversal is not caused by a real physical dipole rotation but by the disappearance of the polar caps and their reappearance with opposite polarity, all taking place at the heliographic poles, and that the tilt of the solar dipole is therefore almost entirely due to the active-region fields. 

In summary, the active regions drive practically all non-axisymmetric fields and high-order fields in the corona. Only the lowest odd-order axisymmetric components, the dipole ($n=1$) and the octupole ($n=3$), follow the polar fields but these components are very influential. Because of their large spatial scale they are very dominant during solar activity minima and are influential during maxima except during polar field reversal. The sunspot number does not represent all active-region fields but the above results indicate a strong correlation between the sunspot number and the strength of high-order ($n\ge 4$) and non-axisymmetric multipoles.

Various explanations for the weakness of the cycle 23 polar fields have been suggested. Via simple numerical estimates and detailed kinematic dynamo modeling, Dikpati~(2011) showed that even a quite modest decrease in active region field strength from one cycle to the next, such as between cycles 22 and 23, could produce a large decrease of polar field strength. Generally speaking, the presence of active region fields is correlated with changes in the polar fields as Figure~\ref{fig:mpoles} shows. However, there are intervals of time when significant quantities of active region field are present in the photosphere but the polar fields do not change significantly. In particular, between 2002 and 2006 there were significant non-axisymmetric (active) fields in the photosphere while the large-scale, odd-order axisymmetric (polar) fields remained remarkably constant. Because there are periods of time when there are active region fields on the Sun that produce no detectable effect on the polar fields, Dikpati's analysis does not fully explain the weakness of the polar fields. In the Babcock-Leighton dynamo model there are two possible explanations for unchanging polar fields in the presence of active regions. According to both explanations, decaying active fields still reach polar latitudes during these time intervals but these decayed active region fields are of such mixed polarity that their net effect on the polar fields is approximately zero. Figure~\ref{fig:butterfly} shows that the plumes of decayed active-region flux moving poleward have been of more mixed polarity since the cycle 23 polar reversal than before. One explanation is that the meridional flows are so fast that the leading polarities in the two hemispheres do not have time to interconnect and interact with each other before being swept poleward (e.g., Schrijver and Liu,~2008; Wang et al.,~2009, Nandy et al.,~2011). To affect the polar fields significantly, these fast meridional flows would have to occur at active latitudes, as Dikpati~(2011) has emphasized. Ulrich's~(2010), Basu and Antia's~(2010) and Hathaway and Rightmire's~(2010) meridional flow speed measurements do not show evidence of significantly faster flows at active latitudes during cycle 23 than during previous cycles.

An alternative explanation is that the active region Joy's law tilts lost their hemispheric bias during cycle 23 (e.g. Jiang et al.~2011, Petrie~2012). In this scenario approximately equal quantities of each polarity would be sent poleward with approximately zero net effect on the polar fields even for slow meridional flow speeds. Petrie~(2012) found that the latitude centroids of the positive and negative active region fields converged in each hemisphere around 2003, implying a disappearing active-region poloidal field around this time. At the same time the high-latitude poleward surges of field were observed to lose their polarity bias in each hemisphere and the polar fields stopped strengthening, consistent with the Babcock-Leighton model. Schrijver and Liu~(2008) and Stenflo and Kosovichev~(2012) have analyzed the the Joy's law tilt angle of selected magnetic bipoles over time and found no statistical change in average Joy's law tilts, and Li and Ulrich~(2012) found from a long-term study that tilt angles of spots appear largely invariant with respect to time at a given latitude, but they decrease slowly during each cycle following the butterfly diagram. On the other hand, Tlatov et al.~(2010) found distinctly different bipole tilt angle distributions for different classes of bipoles during cycles 21-23, and Lef\`evre and Clette~(2011) discovered that the size and complexity distributions of sunspots changed significantly during cycle 23. Some change in the sunspot or bipole size or complexity distributions may be related to the change in the effects of the active region fields on the poles, and the weakness of the polar fields, during cycle 23.

\section{Relation to CME and prominence eruption rates}
\label{sect:cmes}

\begin{figure}[h]
\begin{center}
\resizebox{0.50\hsize}{!}{\includegraphics*{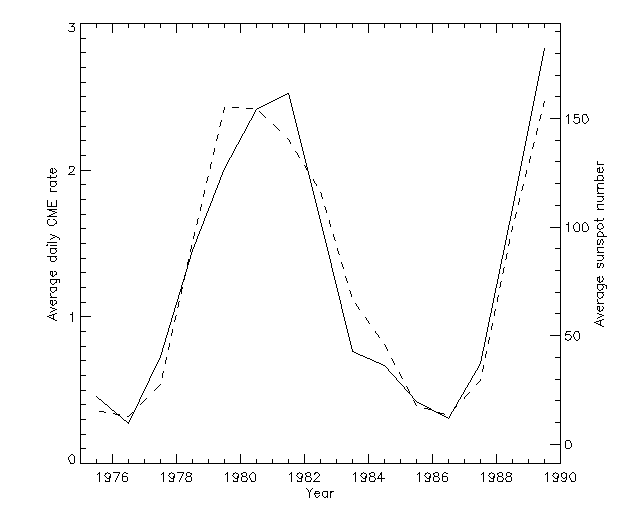}}
\resizebox{0.50\hsize}{!}{\includegraphics*{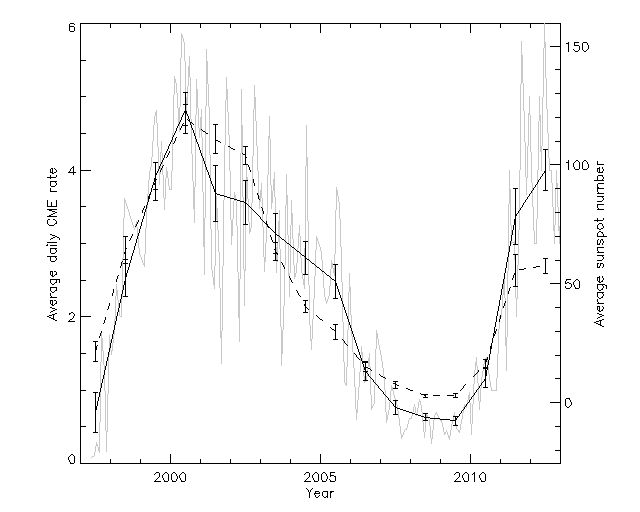}}
\resizebox{0.50\hsize}{!}{\includegraphics*{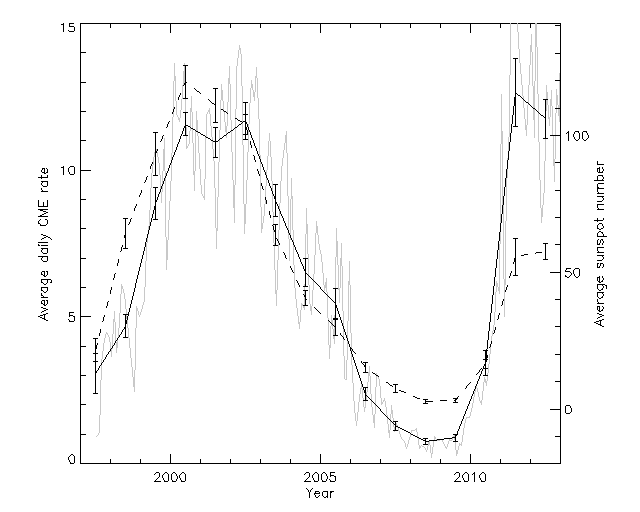}}
\end{center}
\caption{The annual average daily CME rate based on the data collected by Webb and Howard~(1994, top), and the annual (black solid lines) and monthly (grey lines) average daily CME rate estimated by CACTus (middle) and SEEDS (bottom). In all panels the annual averages of the monthly sunspot number are over-plotted (black dashed lines) for comparison. In the middle and bottom plots, standard deviations of the annual means are indicated by error bars.}
\label{fig:cme}
\end{figure}

\begin{figure}[h]
\begin{center}
\resizebox{0.50\hsize}{!}{\includegraphics*{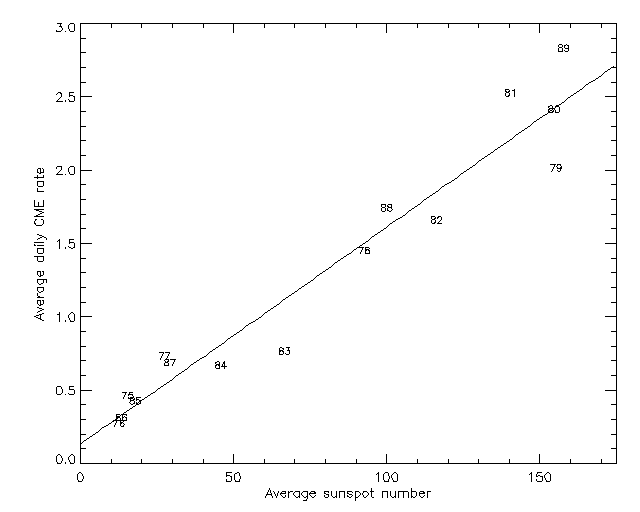}}
\resizebox{0.50\hsize}{!}{\includegraphics*{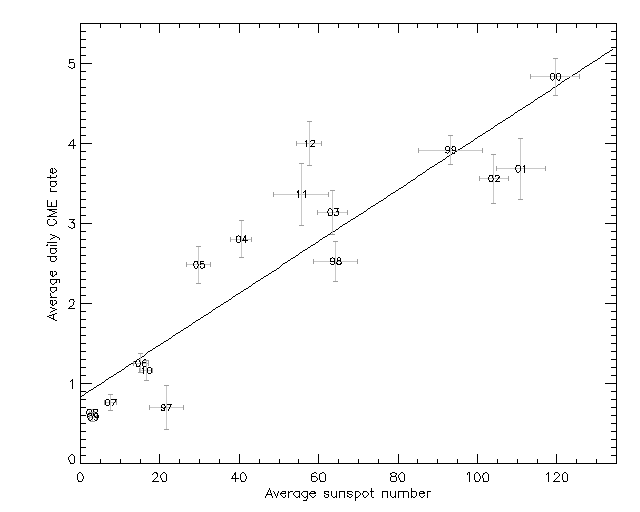}}
\resizebox{0.50\hsize}{!}{\includegraphics*{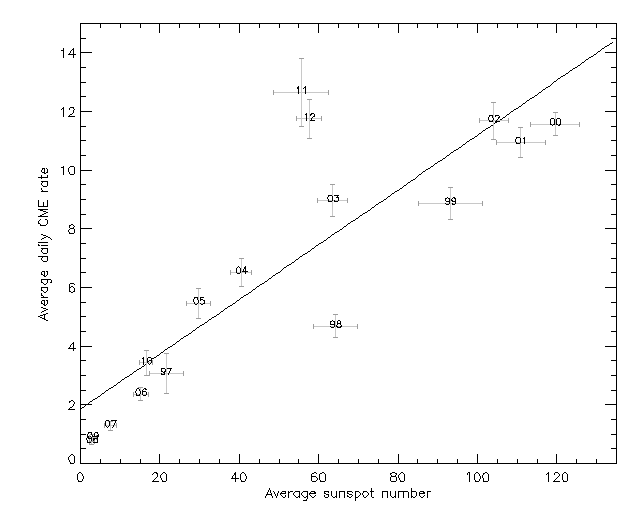}}
\end{center}
\caption{Scatter plots of the annual average daily CME rate against the annual average monthly sunspot number for the based on the data collected by Webb and Howard~(1994,top), and estimated by CACTus (middle) and SEEDS (bottom). The data points are marked by year, 75-89 for the Webb-Howard data and 97-12 for the CACTus and SEEDS data. The linear regression fits are over-plotted. In the middle and bottom plots the error bars indicate standard deviations of the annual means.}
\label{fig:ssncme}
\end{figure}

\begin{figure}[h]
\begin{center}
\resizebox{0.50\hsize}{!}{\includegraphics*{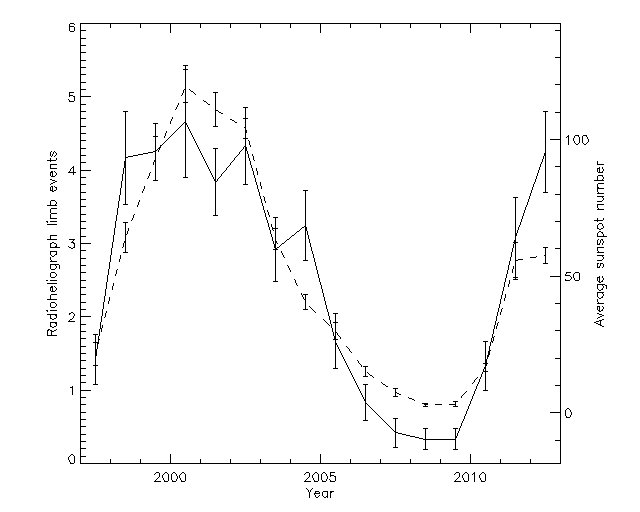}}
\resizebox{0.50\hsize}{!}{\includegraphics*{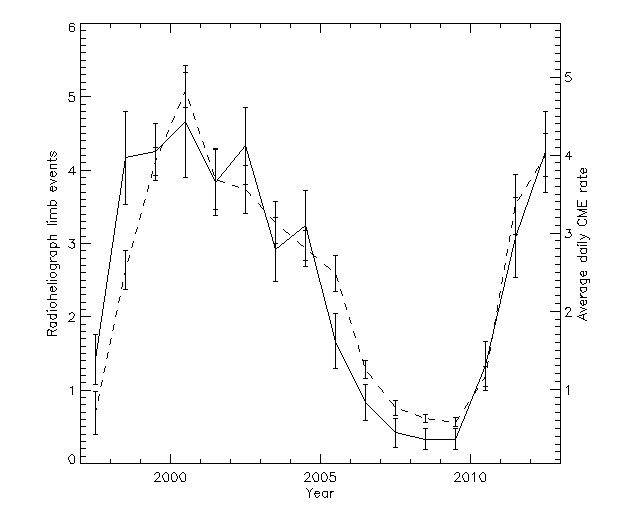}}
\resizebox{0.50\hsize}{!}{\includegraphics*{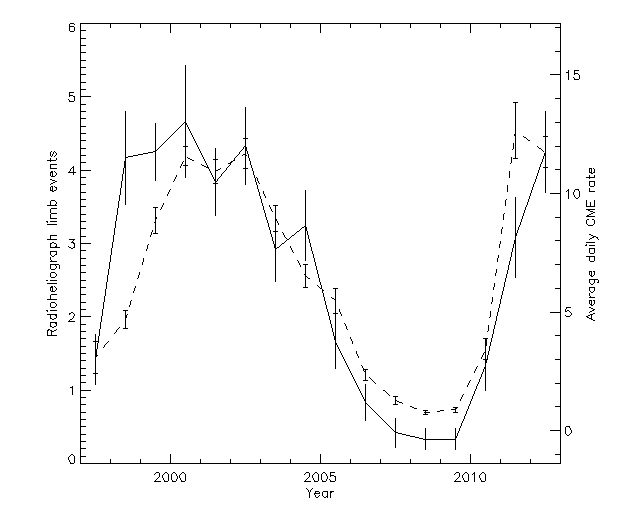}}
\end{center}
\caption{The annual limb prominence eruption counts detected by the Nobeyama Radioheliograph (solid lines), plotted with the average monthly sunspot number (top, dashed line) and the annual average daily CME rate estimated by CACTus (middle, dashed line) and SEEDS (bottom, dashed line). Standard deviations of the annual means are indicated by error bars.}
\label{fig:nobeyama}
\end{figure}

We now investigate consequences of the behavior of the solar field for coronal mass ejection (CME) and prominence eruption (PE) rates. There are two long time intervals over which continuous CME rate data are available. The 15-year period between 1975 and 1989 was studied by Webb and Howard~(1994) who carefully cross-calibrated CME rate information from numerous observational sources. The Royal Observatory of Belgium's Computer Aided CME Tracking (CACTus) project (Robbrecht et al.~2009) has produced estimates for CME rates covering the Large Angle and Spectrometric Coronagraph (LASCO) data archive from 1997 until the present, as has the Solar Eruptive Event Detection System (SEEDS) project of George Mason University. Robbrecht et al.~(2009) presented a detailed study of the CACTus results for cycle 23, comparing these results with those of Webb and Howard~(1994) for cycle 21 and the beginning of cycle 22 (1975-89). They found broad similarities in the behavior of the two data sets, including good correlation with the sunspot number. Here we compare and contrast the data from 1975-1989 compiled by Webb and Howard and the CACTus and SEEDS LASCO data sets covering from 1997 to the end of 2012, with reference to the magnetic results of the previous sections. Luhmann et al.~(2011) has noted that the cycle 24 CME rate has so far been higher than expected in view of the relatively low sunspot number of this cycle. They suggested that this may be related to the weakness of the polar fields during cycle 24, enabling the CME rate to remain comparable to the rates during the two previous stronger field cycles by letting more modest active regions have greater coronal influence.  The tallest closed field lines in the PFSS models shown in Figures~\ref{fig:pfss1} and \ref{fig:pfss2} indicate the boundary between the closed active-region and open coronal-hole fields in the models and they generally correspond to the streamer belts that form above active regions. If the polar fields are strong then these streamer structures would tend to be magnetically stronger and more extensive, and they would therefore more effectively inhibit the eruption of the active fields beneath than if the polar fields were weak. This is why the weakness of polar fields may be related to increased eruption rates.

Figures~\ref{fig:cme} and \ref{fig:ssncme} show that there is excellent correlation between the international sunspot number and the CME rate (Webb and Howard) for the years 1975-1989 and almost as good correlation for 1997-2012. This result confirms past findings that the CME rate follows the solar cycle (Webb and Howard~1994, Robbrecht et al.~2009). As Robbrech et al. note, sunspots only represent a subset of the source regions of CMEs. Several studies have found that the majority of CMEs for which on-disk source regions could be identified are related to filament/prominence eruptions (e.g., Munro et al. 1979; Webb and Hundhausen 1987). However the correlation of these activity indices shows that they are likely to be consequences of a common magnetic activity cycle. Webb and Howard quoted a Pearson correlation coefficient of 0.94 for 1975-1989 but we calculated a coefficient of 0.97 based on a time-weighted average of the various instrument-specific CME rates shown in Webb and Howard's Table~1. The correlation between the sunspot number and the SOHO/LASCO CME rate covering the period 1997-2012 is less impressive, with a Pearson linear correlation coefficient of 0.91 for the CACTus data and 0.85 for the SEEDS data. In particular, the CACTus and SEEDS data sets both tell us that the CME rate for the active years of cycle 24 is unexpectedly high relative to the sunspot number, in line with the comments by Luhmann et al.~(2011). This behavior is more pronounced in the SEEDS data than in the CACTus data. There is no corresponding behavior in the Webb-Howard data for cycle 21.

Clearly there are significant differences between the CME rates plotted in Figures~\ref{fig:cme} and \ref{fig:ssncme}. Besides instrumental differences between the 1975-1989 data and the LASCO data, the manual identification techniques of Webb and Howard differ from the automated algorithms of CACTus and SEEDS, which in turn differ from each other. The SEEDS statistics are two or three times as large as the CACTus statistics for the same LASCO data. We cannot therefore draw conclusions from the amplitudes of the curves in Figure~\ref{fig:cme}. We can, however, compare and contrast the patterns of behavior that these curves represent. It appears that the weaker correlation for 1997-2012 compared to 1975-1989 signifies a fundamental change in the global solar magnetic field than with problems in the CME statistics. Robbrecht et al.~(2009) found that the CACTus statistics for LASCO can be more consistent and more objective than manually compiled statistics - see their Figure~2. The 1997-2012 CME rate statistics follow a simple pattern according to our analysis of two independent catalogs, CACTus and SEEDS, suggesting that a real change of behavior took place. The middle and bottom scatter plots in Figure~\ref{fig:ssncme} show that the active years 2003-06, after the strengthening trend of the polar fields stopped, all lie above the linear regression fit. The two active years of cycle 24, 2011-12, lie further above the regression fit. The CME rates of these two years are comparable with those of the cycle 23 maximum even though the active magnetic fields have been significantly weaker during cycle 24 compared to cycle 23. This change seems to correspond to the polar fields becoming weaker during cycle 23 (Section~\ref{sect:activepolar}). The points below the regression fit consist of the minimum years 2007-09 and the maximum years before the polar fields became weak, 1997-02, with the exception of a year of polar field reversal, 2000, which appears above the regression fit. Thus the active years generally fall below the regression fit before 2003 and above the regression fit after 2003. In the top scatter plot of Figure~\ref{fig:ssncme} for Webb and Howard's 1975-89 data there is no such separation between data points. The weaker correlation between CME rate and sunspot number in the LASCO era may therefore be related to the change in the behavior of the polar fields that occurred around 2002 or 2003.

There are more subtle phase differences between the CME cycles and sunspot cycles. Robbrecht et al.~(2009) reported that during cycle 23 the CME cycle lagged the sunspot cycle by between 6 months and a year, and that the CME cycle rose faster than it declined. They also noted that this behavior may be specific to cycle 23 because Webb and Howard's~(1994) data for cycle 21 did not show it. The middle and bottom panels of Figure~\ref{fig:cme} show that during cycle 23 the CME rate rose, peaked and declined up to a year after the sunspot cycle and that the rise phase was much faster than the declining phase. This pattern agrees with Robbrecht et al.'s~(2009) findings, and is particularly clear in the SEEDS data but is also evident in the CACTus data. The ascent of the CME curves for cycle 24 appear to lead the sunspot number curves but this is because the CACTus and SEEDS CME rates for cycle 24 are so strong compared to the sunspot number. The top panel of Figure~\ref{fig:cme} indicates a quite different pattern: the cycle 21 CME cycle rose more slowly and peaked later than the sunspot cycle, but declined more quickly. The phase differences between the sunspot and CME cycles seem to be complex and variable from cycle to cycle. The change in the rate of eruptions relative to sunspot number that occurred during cycle 23 seems to have been a much simpler change of pattern. 

Further evidence of a real change in the rate of solar eruptions comes in the form of a corresponding pattern in prominence eruption rates (PEs). Figure~\ref{fig:nobeyama} shows the Nobeyama Radioheliograph PE rate compared to the sunspot number and the CACTus and SEEDS LASCO CME rates over the period 1997-2012. Comparing the top panel of Figure~\ref{fig:nobeyama} and the middle and bottom panels of Figure~\ref{fig:cme}, the PE rate has similar behavior to the CME rates: the active years after 2003 tend to lie above the sunspot number curve whereas those before 2003 tend to (but do not all) lie below, while the minimum years lie below. This pattern is not as strong for PEs as for CMEs but it is present in both sets of statistics.  The right panel of Figure~\ref{fig:nobeyama} shows that the CME rate has increased relative to the PE rate since the cycle 23 minimum. The PE rate kept pace with the CME rate during the declining phase of cycle 23 when many polar crown prominences were erupting.

The CME and PE rates may have evolved slightly differently over time because PEs do not occur as high in the atmosphere and therefore may be less inhibited by the overarching streamer structures than CMEs are. Evidence in support of this argument comes from Gopalswamy et al.'s~(2003) result that CME central position angles tend to be offset towards the equator compared to their corresponding PE's, indicating a greater influence of overarching streamer arcade structure on CMEs than on PEs. Alternatively the differences may be related to differences in prominence creation rates over time. Prominences are structures with significant magnetic shear that has either emerged into the atmosphere or been built up there. A relative shortage of prominences may therefore imply a lack of emerged or evolved magnetic shear low in the atmosphere. For example, the presence of overlying closed fields may allow much shear to build up below before an eruption takes place. CMEs, especially the largest ones, are also often associated with shear but CMEs come from a variety of sources. The difference between the CME and PE statistics may therefore also be caused by the weaker polar fields and overarching arcades making it more difficult for atmospheric fields to be sheared into prominence configurations before eruptions take place.

The overall pattern is that eruption rates per sunspot number have generally been higher since the cycle 23 polar field reversal than they were before. This is consistent with Luhmann et al.'s~(2012) interpretation of the polar fields as a restraining influence on eruptions, whose effect has been reduced by the weakening of the polar fields during cycle 23. In Section~\ref{s:multipoles} we found that the polar fields have a dominant influence on the global coronal magnetic field structure over most of the activity cycle via the axial dipole and octupole fields. The statistics from three independent eruption catalogs indicate that the polar fields' influence extends to the eruption rates themselves.

\section{Conclusion}
\label{sect:conclusion}

The PFSS modeling applied in this work represents the simplest approximation to the real global coronal magnetic structure. The PFSS models neglect the effects of forces and electric currents in the low corona and they only crudely model the inertial effects of the expanding solar wind at their outer boundaries. Also these techniques do not fully exploit the full-disk vector spectro-polarimetric magnetic field measurements from the NSO's Synoptic Optical Long-term Investigations of the Sun (SOLIS) and NASA's Helioseismic and Magnetic Imager (HMI) telescopes that have become available in recent years. On the other hand, these standard PFSS techniques make use of the archived scalar magnetic data sets that extend back to the 1970s. They are also well understood and are known to be able to reconstruct the basic global structure of the coronal field cheaply and effectively for reasons explained by Wang and Sheeley~(1992). Such models usually closely approximate much more expensive magnetohydrodynamics (MHD) models in practice (Neugebauer et al.~1998, Riley et al.~2006). For these reasons the PFSS model remains the basic tool for routinely modeling the global coronal field at NOAA's Space Weather Prediction Center\footnote{http://www.swpc.noaa.gov/ws/} and NASA's Community Coordinated Modeling Center\footnote{http://ccmc.gsfc.nasa.gov/models/models\_at\_glance.php}. The decomposition of potential fields into spherical harmonics and multipole components also offers much valuable insight into the evolution of global coronal structure in terms of dominant length scales and symmetries.

Using 3 1/2 solar cycles of NSO/KP and WSO photospheric synoptic data and extrapolated coronal potential field models, we characterized the cyclical evolution of the global photospheric and coronal magnetic fields. The global coronal morphology follows the progress of the magnetic activity cycle, taking a simple dipole-like form during each activity minimum and evolving through series of complex configurations with more structure of smaller active-region scales at the height of each activity maximum. At all phases of the cycle except times of polar field reversal the large-scale, low-order fields, particularly the axisymmetric dipole and octupole, generally dominate the global coronal morphology. The higher-order multipole components of the coronal field, corresponding to the smaller spatial scales, are strongly correlated with the sunspot/active region cycle, as are the non-axisymmetric components of the coronal field. Only the axisymmetric dipole and octupole components correlate well with the polar fields but since these components are generally the largest contributors to the global field, except when the polar field reverses, the polar fields play a determinative role in structuring the global corona. There is no significant signature of polar field reversal in the non-axisymmetric multipoles. The apparent dipole tilt observed over the solar cycle is almost entirely due to the active-region fields. 

Polar field changes are generally well correlated with active fields over most of the period studied, except between 2003-6 when the active fields did not correspond to significant polar field changes. This change in behavior seems to be related to the well-known fact that, the polar fields created during cycle 23 only became about 60\% as strong as those created during the two previous cycles. We see evidence that the correlation between active-region fields and polar fields changes has returned since the cycle 23 minimum and the process of cycle 24 field reversal is well advanced at both poles.

The temporal patterns of the different multipole components reveal the behavior of the global solar field in terms of symmetry and length scale. This kind of analysis does not reveal all properties of the active-region fields, and how these properties compare between this cycle and the previous cycles. There may be different active region size distributions and different distributions of sunspot/active region complexity at different times. These changes are beyond the scope of a multipole analysis like this one but, e.g., Tlatov et al.~(2010) and Lef\`evre and Clette~(2011) have found that significant  changes in such distributions occurred during cycle 23. Also analyses based on longitudinal magnetograms and potential fields cannot inform us of the twist and shear of the fields over activity cycles. But Zhang et al.~(2010) found from more than 20 years of vector magnetograms of active regions taken at observatories in Mees, Huairou and Mitaka that helicity patterns propagate equatorwards following the sunspot cycle but, unlike sunspot polarity, helicity in each solar hemisphere does not change sign from cycle to cycle. What the potential fields did allow us to do was characterize the influence of different classes of photospheric field on the global magnetic field structure, emphasizing the importance of the polar fields.

Statistics from three independent solar eruption catalogs then allowed us to compare the eruption rates to the behavior of the solar magnetic field. Here the influence of the polar fields was again emphasized. The annual average CACTus and SEEDS CME rates based on LASCO data have changed significantly since the cycle 23 polar reversal: they are both systematically higher for active years between 2003-2012 than for those between 1997-2002. The Nobeyama PE rate follows a similar pattern. These results appear to be connected with the weakness of the late-cycle 23 polar fields as suggested by Luhmann, and may be explained by the dominant influence of the polar fields on the global coronal field structure. It would be interesting to investigate the heliospheric consequences of these enhanced eruption rates for the interplanetary magnetic fields solar wind behavior, etc.

This work can be extended using more sophisticated nonlinear force-free and MHD modeling techniques and vector synoptic magnetograms. Such modeling would yield more detailed information on the physical properties of the global coronal field, such as the magnetic helicity and the free magnetic energy, whose spatial and temporal patterns may provide useful insights into the occurrence, size and geo-effectiveness of flares and CMEs and the global behavior of the solar dynamo. 


\acknowledgements{}
I thank an anonymous referee, Sanjay Gosain and Giuliana de Toma for helpful comments on the manuscript. SOLIS data used here are produced cooperatively by NSF/NSO and NASA/LWS. NSO/Kitt Peak 512-channel and SPMG data used here were produced cooperatively by NSF/NOAO, NASA/GSFC, and NOAA/SEL. Wilcox Solar Observatory data used in this study was obtained via the web site http://wso.stanford.edu courtesy of J.T. Hoeksema. This paper uses data from the CACTus CME catalog, generated and maintained by the SIDC at the Royal Observatory of Belgium. The paper also uses CME statistics from the SEEDS project at George Mason University's Space Weather Laboratory, that has been supported by NASA's Living With a Star Program and NASA's Applied Information Systems Research Program. This paper uses data from the Nobeyama radioheliograph at the National Astronomical Observatory of Japan's Nobeyama Radio Observatory.

\end{document}